\documentclass[12pt,preprint]{aastex}

\usepackage{natbib}
\newcommand{\bu}{{\bf u}}

\newcommand{\bg}{{\bf g}}

\newcommand{\be}{{\bf e}}

\shorttitle{Layer formation in double-diffusive convection}
\shortauthors{Rosenblum et al.}

\title{Turbulent mixing and layer formation in double-diffusive convection: 3D numerical simulations
  and theory}

\author{E. Rosenblum$^{1,2}$, P. Garaud$^{2,3}$, A. Traxler$^2$ and S. Stellmach$^4$}
\affil{$^1$ Stony Brook University, Stony Brook, New York, NY-11794, USA \\
$^2$ Department of Applied Mathematics and Statistics, Baskin School of
 Engineering, University of California Santa Cruz, 1156 High Street,
 Santa Cruz, CA-95064, USA \\ 
$^3$ On sabbatical leave at: Institute for Astronomy, 34 `Ohi`a Ku St., Pukalani, HI 96768-8288 \\
$^4$ Institut f\"ur Geophysik, Westf\"alische Wilhelms-Universit\"at
M\"unster, M\"unster D-48149, Germany}

\begin{abstract}
Double-diffusive convection, often referred to as semi-convection in astrophysics, occurs in thermally 
and compositionally stratified systems which are
stable according to the Ledoux-criterion but unstable according to the Schwarzchild criterion. 
This process has been given relatively little attention so far, and its properties 
remain poorly constrained. In this paper, we present and analyze a set of 
three-dimensional simulations of this phenomenon in a 
Cartesian domain under the Boussinesq approximation. We find that in some cases the double-diffusive
convection saturates into a state of homogeneous turbulence, but with 
turbulent fluxes several orders of magnitude smaller than those expected from direct 
overturning convection. In other cases the system rapidly and spontaneously develops
closely-packed thermo-compositional layers, which later successively merge until 
a single layer is left.  We compare
the output of our simulations with an existing theory of layer formation in
the oceanographic context, and find very good agreement between the
model and our results. The thermal and compositional mixing rates increase 
significantly during layer formation, and increase even further with each merger. 
We find that the heat flux through the staircase is a simple function of the layer height.
We conclude by proposing a new approach to studying transport by 
double-diffusive convection in astrophysics.
\end{abstract}

\keywords{convection -- hydrodynamics -- planets and satellites:general -- stars:interior}

\begin{document}

\section{Introduction}

\subsection{Convection, double-diffusive convection (semi-convection) and fingering convection (thermohaline convection)}
 
One of the longest-standing problems in stellar and planetary
astrophysics is that of modeling the transport of heat and chemical 
species within turbulent regions. The best-studied and most
ubiquitously relevant case is that of overturning convection 
through a chemically homogeneous gas layer. There, the well-known Schwarzchild
criterion is used to determine the extent of the convective region,
while the transport properties through the layer are commonly modeled
using mixing-length theory \citep{Biermann1932}. The success
of well-calibrated mixing-length models in explaining many observable 
properties of stars is quite remarkable. 

However, much less is known about convection in the presence of additional factors such as
strong rotation, strong magnetic fields and strong compositional
gradients \citep{spiegel1972}. In all cases, the linear stability
of the system is well-understood \citep{chandrasekhar1961}, 
but characterizing its fully-nonlinear transport properties
remains the subject of ongoing research. 
In this work, we focus on the case of convection in
the presence of a strong stabilizing compositional gradient, but in the
absence of rotation or  magnetic field. This regime is often called ``semi-convection'' in 
astrophysics \citep{schwartzchildharm1958}, 
although we prefer to use the terminology ``double-diffusive convection'' commonly 
used in oceanography to clarify the true
nature of the instability responsible for the turbulence. 

It has long been known that the relevant criterion for instability to overturning convection 
in the presence of a compositional gradient is {\it not} the Schwarzchild criterion,
\begin{equation}
\nabla - \nabla_{\rm ad} = \left( \frac{\partial \ln T}{\partial \ln p} \right ) - \left(
  \frac{\partial \ln T}{\partial \ln p} \right)_{\rm ad} > 0 \mbox{   , }
\end{equation}
but the Ledoux criterion \citep{ledoux1947}: 
\begin{eqnarray}
\nabla - \nabla_{\rm ad}  &>& \nabla_\mu \nonumber \\
\Leftrightarrow \left( \frac{\partial \ln T}{\partial \ln p} \right) - \left(
  \frac{\partial \ln T}{\partial \ln p} \right)_{\rm ad} &>& \left( \frac{\partial \ln \mu}{\partial \ln p} \right) \mbox{   , }
\end{eqnarray}
where $T$ is the temperature, $\mu$ the mean molecular weight, $p$ the pressure, and where the
subscript ``ad'' expresses a derivative at constant specific entropy. 
In fact, both of these criteria merely express the same
property when written in terms of the density stratification: 
\begin{equation}
\left( \frac{\partial \rho}{\partial p} \right)_{\rm ad} > \left( \frac{\partial \rho}{\partial p} \right) \mbox{   . }
\end{equation}
A system is unstable to overturning convection if the
density of a parcel of fluid, raised adiabatically and in pressure
equilibrium from its original position, is lower than that of its new 
surroundings. 

The question of what happens to regions which are stable according to
the Ledoux criterion but unstable according to
the Schwarzchild criterion was first raised by \citet{schwartzchildharm1958}.
It was later found that this regime is in fact {\it also} linearly unstable \citep{walin1964,kato1966}, but
through a {\it double-diffusive} instability, i.e. an instability 
which cannot occur unless the thermal diffusivity of the fluid, $\kappa_T$,
is larger than its compositional diffusivity $\kappa_\mu$. This condition
is however automatically satisfied in stellar and planetary interiors 
where the diffusivity ratio $\tau = \kappa_\mu / \kappa_T$ (often
called the inverse Lewis number), can be as low as $10^{-7}$. 

As a result, a wide range of situations arise in which double-diffusive convection occurs and
controls transport within the object. A commonly studied case is that of semi-convection at the edge of
core-convective stars \citep{ledoux1947,tayler1954,schwartzchildharm1958,merryfield1995}.
In moderately massive stars for example, a mean molecular weight gradient develops over time at the edge of the core, 
and eventually begins to affect convection. When
it is strong enough to stabilize the fluid, the fully-convective region shrinks in size,  leaving behind a
``semi-convective'' region in which transport is
controlled by double-diffusive processes instead. Other related examples are reviewed by \citet{merryfield1995}.

The possible role of double-diffusive convection in regulating thermal and
compositional transport in the interior of giant planets was 
recognized later, and has been discussed in the context of convective
planetary envelopes where the stabilizing component is helium \citep{stevenson1977}, higher-metallicity 
material at the edge of a rocky core (see \citet{stevenson1985} and in particular Figure 2 of
his paper), or methane \citep{gierash1987}. Double-diffusive
convection has also been proposed to explain the abnormally large radii of some transiting
exoplanets \citep{chabrier2007htg}. Finally, it has recently been invoked as a new mechanism for driving
pulsations in white dwarfs \citep{shibahashi2007,kurtz2008}.

Before we move on to describe existing work on double-diffusive convection, 
we note that it should not be confused with that arising from 
the related double-diffusive ``fingering'' instability \citep{stern1960sfa}.  
The latter also occurs in Ledoux-stable systems, but in the opposite 
situation when the more rapidly diffusing thermal field is stably
stratified while the slowly-diffusing compositional field is unstably
stratified. Its turbulent manifestation is often referred to as ``thermohaline convection'' in
astrophysics, by analogy with the oceanic case in which the
compositional gradient is due to salt. We prefer a more general 
terminology and use the alternative name ``fingering convection''. Figure
\ref{fig:regimes} illustrates for clarity the various regimes of
convective instability. See \citet{traxler2010b} for a study of
fingering convection in the astrophysical regime. 

\subsection{Previous work on double-diffusive convection}

Very little is known about mixing by double-diffusive convection, 
despite its obvious importance in stellar and planetary astrophysics.
Linear stability reveals that the unstable modes take the form of
overstable gravity waves \citep{walin1964,kato1966,baines1969}.  What governs the saturation 
of the instability in the astrophysically-relevant parameter regime, however, remains
essentially unknown. Note that by ``astrophysically-relevant'' we imply a low diffusivity 
ratio, $\tau \ll 1$, and a low Prandtl number, ${\rm Pr} = \nu/\kappa_T \ll 1$, where $\nu$ is the viscosity. 
Both numbers are typically of the order of $10^{-5} - 10^{-7}$ in stellar and planetary interiors.

To add to the complexity of the problem, double-diffusive convection is known in some cases 
to lead to thermo-compositional layering, i.e. to the development of stacks of 
well-mixed fully-convective layers separated by strongly stratified interfaces. This double-diffusive 
layering is commonly observed for example in the arctic ocean
\citep{neal1969,toole2006} where cool fresh water lies on top of
warmer, saltier water. It has been studied extensively in 
laboratory experiments \citep{turner1968,huppert1979,huppert1980,turner1985mc}.
An important result of these studies is that turbulent mixing in the presence of
layers is significantly enhanced compared with that of a
system which has the same {\it overall} contrast in temperature and
composition, but where the stratification is everywhere much smoother.  

Whether layer formation occurs in double-diffusive convection at low Prandtl number actually
remains to be determined -- it is usually {\it assumed}
\citep{spruit1992,chabrier2007htg}, by analogy with the high-Prandtl number
oceanographic case. It is important to realize, however, that such analogies can be
misleading. This was recently demonstrated by \citet{traxler2010b} in the case of fingering convection.
Similar thermohaline staircases are ubiquitously
observed in {\it fingering-unstable} regions of the ocean \citep{schmitt2005edm}, and
have been shown to form
spontaneously through a secondary linear instability of homogeneous fingering convection
\citep{radko2003mlf,traxler2010,stellmach2010} (see
\S\ref{sec:layers} too). However, \citet{traxler2010b} demonstrated
that this secondary instability does not happen in the astrophysical
context and concluded that thermo-compositional layers are not expected in that
case. In other words, given that the analogy with the heat-salt system doesn't hold in the fingering regime, 
one should be extremely cautious about using it {\it a priori} in the double-diffusive regime. 

To summarize, it is known that ``homogeneous'' and ``layered'' double-diffusive convection have 
rather different mixing properties. A good mixing parametrization therefore needs
to incorporate both cases, and must include a criterion to decide whether the system considered lies in
one or the other. Existing parametrizations, by contrast, have so far either
ignored the possible effect of layering \citep{schwartzchildharm1958,langer1983}, or relied on
it \citep{spruit1992}. 

A few numerical simulations of double-diffusive  convection have been performed to
date to address the problem. The first of this kind (to our knowledge), were presented by
\citet{merryfield1995}. He ran a series of two-dimensional (2D) anelastic simulations,
horizontally-periodic, and bounded in
the vertical direction by two plates. Double-diffusive convection was forced 
through the imposed boundary conditions,
which maintained a given overall temperature
and compositional contrast across the domain. \citet{merryfield1995}
focused on understanding how the mechanism responsible for the
saturation of the initial double-diffusive instability depends on the governing parameters, and
in particular the strength of the thermal driving, the Prandtl number
and the diffusivity ratio. He also compared the outcome of his simulations
with existing parametrizations of double-diffusive convection both
in the absence of layers, and in the presence of initially forced layers.
One of the main difficulties encountered was the
development of numerical instabilities in cases with strong thermal driving, which
prevented him from drawing definite conclusions on the long-term
statistical properties of the turbulence. In addition, in many of the runs
the flows were eventually influenced by the presence of domain boundaries. 

In subsequent years, two additional attempts at modeling
double-diffusive convection were made.
\citet{biello2001}, as part of his PhD thesis, ran a series of 2D
fully-compressible simulations which complement
those of \citet{merryfield1995}. There, the system was also confined
between two plates, but the boundary conditions were ``fixed flux''
conditions. \citet{biello2001} was interested in 
studying more specifically the layer formation process, and his experimental setup was
similar to that of heat-salt laboratory experiments \citep[e.g.][]{huppert1979}. For this purpose,
the simulations were initialized with {\it stable} uniform gradients in temperature and 
composition, but destabilized at $t=0$ by increasing the
heat flux at the bottom boundary. He found that the first bottom
layer easily forms, but did not observe any subsequent layer
formation. He analyzed the dynamics of the interface, and
concluded that interfacial transport was dominated by wave-breaking 
rather than by diffusive processes as is often assumed. However, boundary effects also began 
to influence the results of his simulations after some time. \citet{bascoul2007}, also 
as part of his PhD thesis, studied a similar 2D time-dependent system,
in which the initial background state had a homogeneous composition
and neutrally stable temperature gradient, and where the system was
destabilized by an imposed heat and mean-molecular weight flux through
the bottom boundary. He also observed the growth of a 
convective layer near the bottom boundary and studied its
development, for high Prandtl number (heat-salt regime) and for low
Prandtl number (``astrophysical'' regime). He was unable, however, to run his simulations 
long enough to achieve statistical equilibrium. 

In this paper, we present a new series of three-dimensional (3D) numerical experiments to study mixing 
by double-diffusive convection. We approach the problem from a different 
but complementary angle, and try to address some of the inherent shortcomings of the experimental setup 
used in previous studies: we focus {\it specifically} on measuring the quasi-steady 
statistical properties of double-diffusive turbulence,  and use a numerical 
setup which minimizes the effects of domain boundaries. 

We discuss the model setup and briefly summarize its linear stability
properties in \S\ref{sec:model}. The numerical algorithm and the
selection of the experimental parameters are described in
\S\ref{sec:num}. In each case, as 
described in \S\ref{sec:homogen}, we extract values for the transport
coefficients while the system is in a state of homogeneous turbulence.  However,
we find that for more unstable systems a secondary instability leads to the formation
of thermo-compositional layers. The layers continue evolving and
successively merge, and each merger is accompanied with a significant increase in the
transport coefficients.  These results are discussed in \S\ref{sec:layers}, and compared with a
recent theory of layer formation in the oceanographic context \citep{radko2003mlf}. We find
good agreement between this theory and our numerical results, which enables us to deduce a general
criterion for the spontaneous formation of layers in double-diffusive convection in the
astrophysical context. We discuss our results in \S\ref{sec:ccl}.

\section{Model description and linear analysis}
\label{sec:model}

In this section we present the governing equations and briefly
summarize known results on the linear stability of the problem. 
Our formalism is overall very similar to the one used by \citet{traxler2010b} for fingering convection. 

\subsection{Governing equations and boundary conditions}
\label{sec:goveqs}

As we demonstrate later (see \S\ref{sec:linear}), the typical lengthscale of the 
unstable motions (in the absence of layers) is of the order of a 
few kilometers at most in parameter regimes typical of stellar and planetary interiors. 
It is therefore justified to neglect the effect of curvature entirely and work in a local 
Cartesian reference frame $(x,y,z)$. Here, gravity defines the vertical direction: $\bg = -g \be_z$. 
We ignore the possible presence of magnetic fields for simplicity, and neglect the effect of rotation.
The latter is justified whenever the mean rotation rate is much smaller than the buoyancy (Brunt-V\"ais\"al\"a) frequency, which is often the case. 

In all that follows, we simplify the problem by using the Boussinesq approximation
\citep{spiegelveronis1960}. This approximation is a regular 
asymptotic limit of the primitive governing equations (for mass, momentum and energy conservation) 
when the domain height $L_z$  is much smaller than the density scaleheight $D_\rho$ (i.e. $L_z /D_\rho \rightarrow  0)$, and when the typical flow velocity $u$ is much smaller than the sound speed $c_s$ (i.e. $u/c_s \rightarrow 0$). 
It is therefore particularly relevant here since the typical lengthscale of the perturbations is much smaller than a pressure or density scaleheight  (unless the region considered is {\it very} close to the photosphere), and the typical velocities are always substantially subsonic (unless the system is {\it very} close to the onset of overturning convection). 

Since we consider a small fluid region within a much larger system, 
we can assume that the background composition and temperature gradients 
$\mu_{0z}$, $T_{0z}$ and $T^{\rm ad}_{0z}$ are constant. Here, the index $0$ denotes a background field, and the index $z$ denotes a derivative with respect  to the vertical coordinate $z$. 
We restrict our study to the case of double-diffusive convection by choosing the background stratification 
such that $\nabla_\mu > \nabla - \nabla_{\rm ad} >0$, or equivalently $\mu_{0z} < T_{0z} - T_{0z}^{\rm ad} < 0$. 

In the Boussinesq approximation the mass conservation equation is replaced by the continuity equation 
\begin{equation}
\nabla \cdot \bu = 0 \mbox{   , }
\end{equation}
where $\bu = (u,v,w)$ is the velocity field,  and the (dimensional)
thermal energy equation is approximated \citep[e.g.][]{ulrich1972} by
\begin{equation}
\frac{\partial T}{\partial t} + \bu \cdot \nabla T  + (T_{0z} - T^{\rm
  ad}_{0z}) w =     \kappa_T \nabla^2 T \mbox{   , }
\end{equation}
where $T$ now represents the dimensional temperature
perturbation (all background quantities being denoted by the subscript
$0$ instead), and we have assumed 
for simplicity that the thermal diffusivity is constant. 
This equation is derived from an energy conservation principle, 
noting that in the Boussinesq approximation temperature and entropy perturbations are proportional.
The term $T_{0z} - T^{\rm ad}_{0z}$ therefore models the advection of the background {\it entropy}
 gradient. 

In the Boussinesq approximation, the density, temperature and mean molecular weight ($\rho$, $T$ and $\mu$ respectively) are related via
\begin{equation}
\frac{\rho}{\rho_0}  = - \alpha T + \beta \mu  \mbox{   , }
\end{equation}
where $\rho_0$ is the mean density of the region considered, and $\alpha$ and $\beta$ are the coefficients of 
thermal expansion and compositional contraction respectively 
(e.g. $\alpha = 1/T_0 $ and $\beta = 1/\mu_0$ for a perfect gas, where $T_0$ and $\mu_0$ are the mean 
temperature and mean-molecular weight in the region considered). 

We construct our numerical model in such a way as to minimize the effects of the computational domain boundaries. For this purpose we use a triply-periodic box of size $(L_x,L_y,L_z)$, in which convection is permanently forced by the aforementioned background stratification. This approach has recently been used 
with success in modeling and studying the formation of thermohaline staircases in the oceanographic context by \citet{radko2003mlf} and \citet{stellmach2010}, and is discussed in more detail in these papers. 
In this framework, the temperature and mean molecular weight fields can be written as the sum of a background variation plus perturbations which are triply-periodic functions of $(x,y,z)$ such that 
\begin{equation}
T(x,y,z,t) = T(x+L_x,y,z,t) =T(x,y+L_y,z,t) = T(x,y,z+L_z,t) \mbox{   , }
\end{equation}
and similarly for $\mu$. The pressure perturbation $p$  and velocity field $\bu$ are also assumed to be 
triply-periodic functions in the same way.

We non-dimensionalize the equations using the anticipated lengthscale of the fastest growing modes of linear instability, which is a thermal diffusion scale \citep[e.g.][]{baines1969}: 
\begin{equation}
d = \left( \frac{\kappa_T \nu}{\alpha g |T_{0z} - T^{\rm ad}_{0z} |} \right)^{1/4} = \left( \frac{\kappa_T \nu}{N^2} \right)^{1/4} \mbox{   , }
\end{equation}
where $N$ is the buoyancy frequency. Very roughly, in typical stellar interiors, $\nu = O(10)$cm$^2$/s, $\kappa_T = O(10^7)$cm$^2$/s and $N^2 = O(10^{-6})$s$^{-2}$ so $d = O(10^{3.5})$cm, or in other words, {\it a few hundreds of meters} only.  

Note that with this definition, the thermal Rayleigh number defined on the finger scale is exactly one, while the 
global Rayleigh number,
\begin{equation}
{\rm Ra}_T = \frac{\alpha g |T_{0z} - T^{\rm ad}_{0z}| L_z^4}{\kappa_T \nu}  = \left(\frac{L_z}{d} \right)^4
\end{equation}
is a function of the dimensionless height of the domain {\it only}.
The unit timescale is taken to be the diffusion timescale across $d$,
namely $[t] = d^2/\kappa_T$, and the velocity scale is $[v]
= \kappa_T/d$. The unit temperature is $[T] = d |T_{0z} -
T^{\rm ad}_{0z}|$, and the unit mean-molecular weight is $[\mu] =
(\alpha/\beta) |T_{0z} -T^{\rm ad}_{0z}|  d $. The resulting non-dimensional governing equations are 
\begin{eqnarray}
\frac{1}{{\rm Pr}}\left(\frac{\partial \tilde \bu}{\partial t} +
  \tilde\bu \cdot \nabla  \tilde\bu\right) &=& -\nabla \tilde{p}
  + (\tilde{T}-\tilde{\mu}) \be_z +\nabla^2 \tilde\bu \mbox{   , } \nonumber \\
\frac{\partial \tilde{T}}{\partial t} +  \tilde\bu \cdot \nabla
\tilde{T} -  \tilde w &=& \nabla^2 \tilde{T} \mbox{   , }\nonumber \\
\frac{\partial \tilde{\mu}}{\partial t} +  \tilde\bu \cdot \nabla
\tilde{\mu} - R_0^{-1}  \tilde w  &=& \tau \nabla^2 \tilde{\mu} \mbox{   , }\nonumber \\
\nabla \cdot  \tilde\bu &=& 0 \mbox{   , }
\label{eq:goveqs}
\end{eqnarray} 
where quantities with tildes now represent the dimensionless,
triply-periodic perturbations. 
The non-dimensionalization introduces three parameters, namely the
aforementioned 
Prandtl number Pr$=\nu/\kappa_T$  and diffusivity ratio
$\tau=\kappa_\mu/\kappa_T$, as well as the so-called {\it density ratio} 
\begin{equation}
R_0 = \frac{\alpha |T_{0z} - T_{0z}^{\rm ad}| }{ \beta |\mu_{0z}| }= \frac {\nabla - \nabla_{\rm ad}}{\nabla_\mu} \mbox{   . }
\label{eq:R0def}
\end{equation}
Finally, note that for reasons described in \S\ref{sec:R0-1}, it is
common and preferable to work with the inverse density ratio
$R_0^{-1} = 1/R_0$ as a governing parameter. 

\subsection{Linear stability}
\label{sec:linear}

The linear stability of this problem is well-understood, thanks to the
works of \citet{walin1964}, \citet{veronis1965} and \citet{baines1969}
in the oceanographic context, and \citet{kato1966} in the
astrophysical context. The salient points are summarized here for 
completeness and clarity, as some of them will be used later.

To analyze the stability of the governing equations we first
linearize them around $\tilde{T}=\tilde{\mu}=0$ and
$\tilde \bu=0$, and assume normal forms for all perturbations as $q = \hat q e^{ilx + imy + ikz + \lambda t}$ where hatted quantities are the mode amplitudes, $l$, $m$, and $k$ are horizontal and vertical wavenumbers respectively, and $ \lambda $ is the growth rate. This procedure yields a cubic equation for $\lambda$ in terms of the wavevector ($l$,$m$,$k$) and of the non-dimensional parameters:
\begin{equation}
\left(\frac{\lambda}{{\rm Pr}} + K^2\right)(\lambda + K^2)(\lambda + \tau
K^2)\left(\frac{K^2}{l^2+m^2}\right) - (\lambda + \tau K^2) +R_0^{-1} (\lambda + K^2) = 0 \mbox{   , }
\label{eq:fastest}
\end{equation}
where $K^2 = l^2+m^2+k^2$. 

It can be shown that the most unstable mode always occurs when the
vertical wavenumber, $k$, is equal to zero. This corresponds to an
"elevator mode", similar to the ones found in the related problem of
fingering convection and of homogeneous Rayleigh-B\'enard
convection. Furthermore, without loss of generality, we may select $l$
or $m$ to be zero by reorienting the Cartesian frame so that the
fastest growing modes can be described using one horizontal wavenumber 
only. The cubic equation (\ref{eq:fastest}) then becomes:
\begin{equation}
\left(\frac{\lambda}{{\rm Pr}} + l^2\right)(\lambda + l^2)(\lambda + \tau l^2) -
(\lambda + \tau l^2) + R_0^{-1} (\lambda + l^2) = 0 \mbox{   . }
\label{eq:fastest2}
\end{equation}
There are typically one real and two complex roots of this
equation. The real root is always negative, achieving a maximum of
zero when $l=0$. However, the complex root yields a
positive growth rate when the inverse density ratio lies in the interval $R_0^{-1} \in \left( 1 , \frac{{\rm Pr}+1}{{\rm Pr}+\tau} \right) $. The oscillation frequency is close to the buoyancy frequency. The unstable modes can therefore be viewed as overstable gravity waves, as mentioned earlier. 

\begin{figure}[h]
\epsscale{0.7}
\plotone{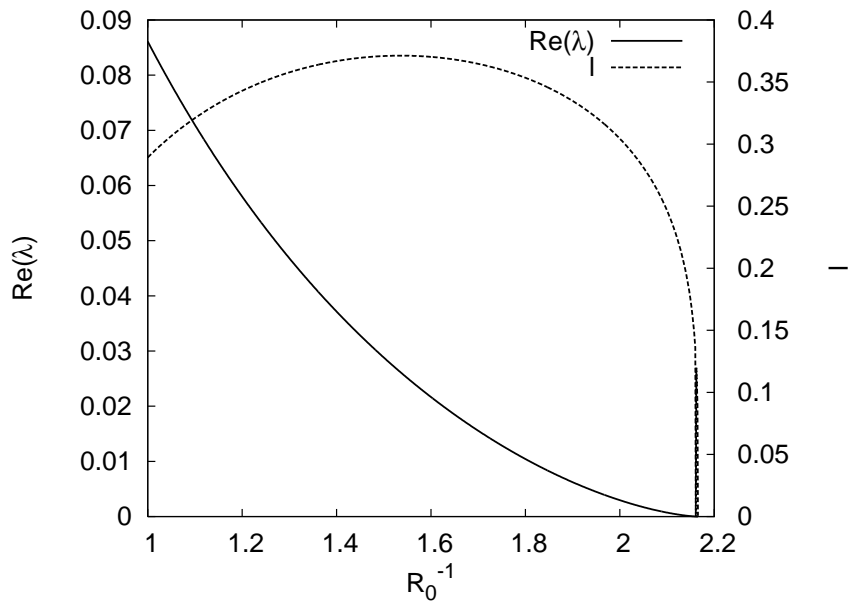}
\caption{Growth rate (solid line) and horizontal wavenumber $l$ (dotted line) of the
  most rapidly growing gravity wave (i.e. the double-diffusive mode of instability), as a function of the inverse density
  ratio, for ${\rm Pr}= \tau = 1/3$. All quantities plotted are in the units used in this paper
  (see \S\ref{sec:goveqs}).} 
\label{fig:mostunstable}
\end{figure}

By maximizing the real part of $\lambda$ over all horizontal
wavenumbers, we can find the most unstable mode in the
system. Figure \ref{fig:mostunstable} shows its wavenumber and 
growth rate for ${\rm Pr}=\tau=1/3$. By and
large, the most rapidly growing mode has wavelength of the order of
20$d$ regardless of $R_0^{-1}$ for the parameters selected, which implies a lengthscale of a few kilometers
as described in \S\ref{sec:goveqs}. 

\subsection{$R_0$ vs. $R_0^{-1}$}
\label{sec:R0-1}

\begin{figure}[h]
\epsscale{0.7}
\plotone{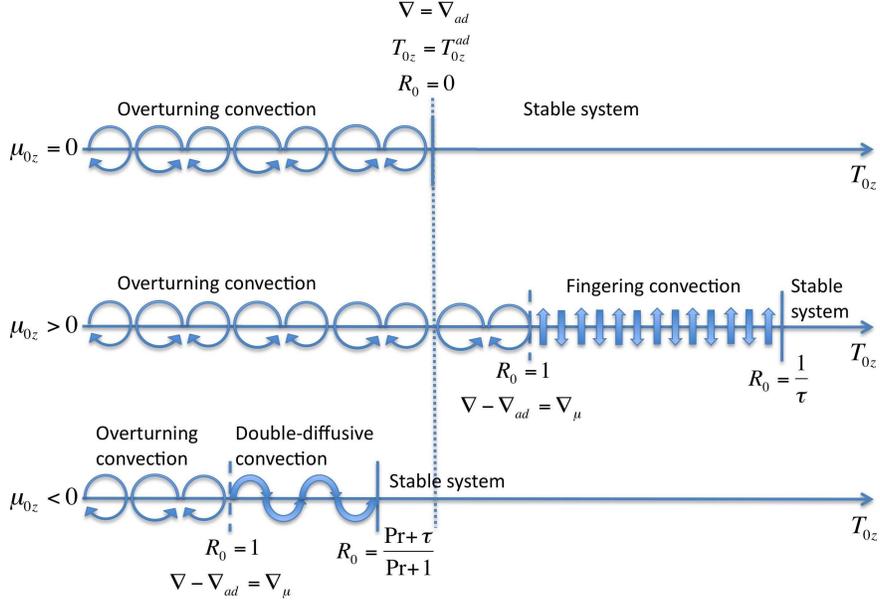}
\caption{Summary of the various regimes of convective instability. The
top line corresponds to the chemically homogeneous case, with
$\nabla_\mu = 0$. The system is unstable to
direct overturning convection if $\nabla > \nabla_{\rm ad}$ or
equivalently $T_{0z} < T_{0z}^{\rm ad}<0$. The Schwarzchild criterion
(dotted line) appropriately marks the stability boundary (solid line). In the presence of an
unstable mean molecular weight gradient (middle line), the region unstable to convective
overturning extends into the subadiabatic regime, and is stabilized
only when $\nabla-\nabla_{\rm ad} = \nabla_\mu$ (Ledoux
criterion, vertical dashed line). The Schwarzchild criterion in this
regime is not relevant. Beyond the Ledoux limit, the system can {\it still} be unstable, this time
to fingering convection. In the case where the system has a stable
mean molecular weight gradient (bottom line), the region of parameter space
unstable to overturning convection shrinks according to the Ledoux
criterion. The system can {\it still} be unstable to
double-diffusive convection in a subset of the interval between
Ledoux-stability and Schwarzchild stability.  } 
\label{fig:regimes}
\end{figure}

As mentioned previously, it is standard in oceanographic studies of
double-diffusive convection to use the parameter $R_0^{-1} = \beta \mu_{0z} / \alpha T_{0z}$, while
$R_0 = \alpha T_{0z} / \beta \mu_{0z}$ is used in studies of fingering convection\footnote{For added
  confusion, studies of double-diffusive convection often call $\beta \mu_{0z} / \alpha T_{0z}$ the density ratio and denote it as $R_0$ although we will not use this naming convention here.}. The reason for this
change in convention is to emphasize the symmetry between the two
regimes, which is apparent in Figure \ref{fig:regimes}. Indeed, in
that case, one can write that overturning convection 
occurs for 
\begin{eqnarray}
&& R_0 < 1  \mbox{   fingering regime, } \nonumber \\
&& R_0^{-1} < 1 \mbox{   diffusive regime, }
\end{eqnarray}
the double-diffusive regime occurs for
\begin{eqnarray}
&& 1 \le R_0 \le \frac{1}{\tau}  \mbox{   fingering regime, } \nonumber \\
&& 1 \le R_0^{-1} \le \frac{{\rm Pr} + 1}{{\rm Pr} + \tau}  \mbox{   diffusive regime, }
\end{eqnarray}
and finally that the system is stable for 
\begin{eqnarray}
&&  R_0 > \frac{1}{\tau}  \mbox{   fingering regime, } \nonumber \\
&& R_0^{-1} > \frac{{\rm Pr} + 1}{{\rm Pr} + \tau}  \mbox{   diffusive regime. }
\end{eqnarray}
The symmetry is even more striking at low Prandtl number, where
the critical value for complete stability is much larger than one 
in both cases. For the diffusive regime, this implies that
the system is fully convective up to the Ledoux-stability criterion,
and then double-diffusively convective between Ledoux-stability and
Schwarzchild-stability. 
A summary of the various regimes of instability, for the
compositionally homogeneous case, for the fingering case and for this
diffusive case is presented in Figure \ref{fig:regimes}. 

\section{Numerical Experiments}
\label{sec:num}

\subsection{Description of the experiments}
\label{sec:numexp}

We solve the governing equations and boundary conditions
presented in \S\ref{sec:goveqs} using a high
performance spectral code developed by S. Stellmach
\citep{traxler2010,stellmach2010,traxler2010b}, specifically designed
for the study of fingering convection in the oceanographic
context. The code can be used ``as is'' to model double-diffusive
convection simply by reversing the sign of both background
gradients. Since it was developed for studying oceanic
convection, the code does not consider entropy separately from
temperature. As a result, it intrinsically assumes that $T_{0z}^{\rm
  ad} = 0$, so that $R_0 = \alpha T_{0z} / \beta \mu_{0z}$. 
By virtue of being non-dimensional, however, the results can nevertheless
straightforwardly be applied to the astrophysical context simply by
{\it interpreting} $R_0$ as being defined by (\ref{eq:R0def}). 
 
Our goals are threefold: (a) to characterize transport by homogeneous double-diffusive convection (i.e. in the absence of layers), (b) to determine if, under which conditions, and through which process thermo-compositional layers may form and (c) to characterize transport by layered double-diffusive convection if appropriate. 

For this purpose, we ran a sequence of exploratory numerical experiments setting ${\rm Pr} = 1/3$ and $\tau = 1/3$. We selected ${\rm Pr}$ and $\tau$ below unity to be in the ``low Prandtl number, low diffusivity ratio'' regime, but not too small so that we could remain in a numerically-tractable region of parameter space. While it is possible to run simulations for lower values of these parameters in 3D \citep[see][for example]{traxler2010b}, these are computationally {\it much} more demanding. In this first analysis, we wanted to be able to run a uniform set of simulations across the whole instability range, and integrate some of them for a significant length of time to observe the layer formation and merger process if and when it occurs.  We discuss the applicability of our results to lower Prandtl number and lower diffusivity ratio environments in \S\ref{sec:ccl}.

In order to get statistically meaningful measurements of the turbulent
fluxes in this system, to address points (a) and (c) raised above,  
we must ensure that the computational box contains at least a few wavelengths of
the most unstable mode in the horizontal directions \citep{traxler2010}. 
As found in Figure \ref{fig:mostunstable}, these are of the order of about $20d$ so we
select a domain size with $L_x = L_y = 100d$ in all simulations. We
use an aspect ratio of 1 and choose $L_z = 100$, which corresponds to
${\rm Ra}_T = 10^8$. For comparison, we also ran a series of
simulations with narrower domains ($L_x = L_y = 50d$) at ${\rm Ra}_T =
10^8$ for all $R_0^{-1}$, as well as one taller-domain simulations using $L_z = 178d$
(equivalently, ${\rm Ra}_T = 10^9$) for $R^{-1}_0 = 1.2$. See Table 1 for a comparison of these runs.
In dimensional terms, the domain height is of the order 
of a few tens of kilometers using the estimates presented in \S\ref{sec:goveqs} for typical stellar interiors, which is {\it much} smaller than a pressure or density scaleheight and therefore fully justifies the use of the Boussinesq approximation. 

In terms of spatial resolution, we use a
sufficient number of Fourier modes in all simulations to resolve the typical size of the
composition and velocity perturbations (which are roughly of the same
size when $\tau = {\rm Pr}$). Similar runs with different spatial resolutions 
are presented in Table 1, and show consistent results. 


\begin{table}
\begin{center}
\caption{Summary of runs performed and measured turbulent fluxes in the homogeneous phase. }
\begin{tabular}{cccccccc}
\\
\tableline
$R_0^{-1}$ & $N_x,N_y$ & $N_z$ & $t_{\rm start}$ & $t_{\rm end}$ & ${\rm Nu}_T$ & ${\rm Nu}_\mu$ &$1/\gamma_1$  \\
1.1 & 192 & 192 & 260 & 370 & 5.7$\pm$3.9 & 11.2$\pm$4.2 & .63$\pm$.10 \\
1.1$^{(a)}$   & 192 & 192 & 270 & 460 & 5.0$\pm$1.2 & 9.6$\pm$3.5 & .61$\pm$.10 \\
1.15 & 192  & 192 & 300 & 630 & 3.9$\pm$.7 & 6.7$\pm$2.3 &.58$\pm$.11 \\
1.15$^{(a)}$ & 96  & 192 & 290 & 510 & 3.7$\pm$.7 & 6.3$\pm$2.2 &.57$\pm$.11 \\
1.2  & 96  & 96  & 300 & 1000 & 3.4$\pm$.6 & 5.5$\pm$1.8 & .57$\pm$.10 \\
1.2$^{(a)}$  & 96  & 96  & 310 & 930 & 2.6$\pm$.5 & 4.0$\pm$1.6 & .52$\pm$.13 \\
1.2$^{(b)}$ & 96 & 192 & 316 & 791 & 2.6$\pm$.3 & 3.9$\pm$.9 & .54$\pm$.07  \\
1.2$^{(c)}$ & 96 & 192 & 316 & 1044 & 3.1$\pm$.5 & 5.0$\pm$1.7 & .56$\pm$.11  \\
1.2$^{(c)}$ & 192 & 384 & 316 & 885 & 3.0$\pm$.4 & 4.7$\pm$1.3 & .57$\pm$.09  \\
1.35 & 96  & 96  & 420 & 900 & 1.9$\pm$.3 & 2.5$\pm$.8 & .51$\pm$.11 \\
1.35$^{(a)}$ & 96  & 96  & 430 & 1320 & 1.8$\pm$.3 & 2.2$\pm$.9 & .49$\pm$.15 \\
1.5  & 96  & 96  & 650 & 1450 & 1.5$\pm$.2 & 1.8$\pm$.5 & .51$\pm$.09 \\
1.5$^{(a)}$  & 96  & 96  & 600 & 1530 & 1.5$\pm$.2 & 1.7$\pm$.8 & .49$\pm$.17  \\
1.6  & 96  & 96  & 800 & 1100 & 1.4$\pm$.1 & 1.5$\pm$.3 & .52$\pm$.07 \\
1.6$^{(a)}$  & 96  & 96  & 750 & 880 & 1.4$\pm$.1 & 1.5$\pm$.4 & .52$\pm$.09 \\
1.85 & 48  & 96  & 2000 & 2500 & 1.17$\pm$.03 & 1.2$\pm$.1 & .57$\pm$.04  \\
1.85$^{(a)}$ & 48  & 96  & 1940 & 2500 & 1.19$\pm$.05 & 1.2$\pm$.2 & .57$\pm$.05  \\
2.1  & 48  & 96  & 0 & 2500 & 1 & 1 &.6306   \\
\tableline
\end{tabular}
\tablecomments{For each value of the inverse density ratio $R_0^{-1}$,
  the second and third columns show the horizontal and vertical
  number of grid-points used. The times $t_{\rm start}$ and $t_{\rm end}$ define
  the interval over which the turbulent fluxes, measured
  via ${\rm Nu}_T$ and ${\rm Nu}_{\mu}$, were averaged. The quantity $\gamma_1$
  is the primary way of measuring the total flux
  ratio $\gamma^{\rm tot}$; see text for detail. Errors quoted indicate the rms fluctuations of ${\rm Nu}_T(t)$, ${\rm Nu}_{\mu}(t)$ and $1/\gamma_1(t)$ about the respective means.  Unless otherwise
  specified (see footnotes), simulations were run at domain sizes of
  $100d \times 100d \times 100 d$. Note that the simulation for $R_0^{-1} = 2.1$ was integrated for significant time, but by the end of the run the perturbations had not grown to high-enough amplitudes to induce a statistically significant heat or compositional flux.}
\tablenotetext{a}{$50d \times 50d \times 100d $}
\tablenotetext{b}{$44d \times 44d \times 178 d $}
\tablenotetext{c}{$89d \times 89d \times 178 d $}
\end{center}
\end{table}

\subsection{Qualitative description of the results}
\label{sec:numdesc}

We ran a first set of cubic-domain simulations ($L_x=L_y=L_z =100d$) for $R_0^{-1}$ varying across the whole instability range, which for our selected parameters corresponds to $1<R_0^{-1}<2.167$.
We found that the initial behavior of the system is qualitatively similar for all density ratios: the perturbations first grow exponentially, and then saturate into a homogeneous turbulent state. 

\begin{figure}[h!]
\epsscale{0.7}
\plotone{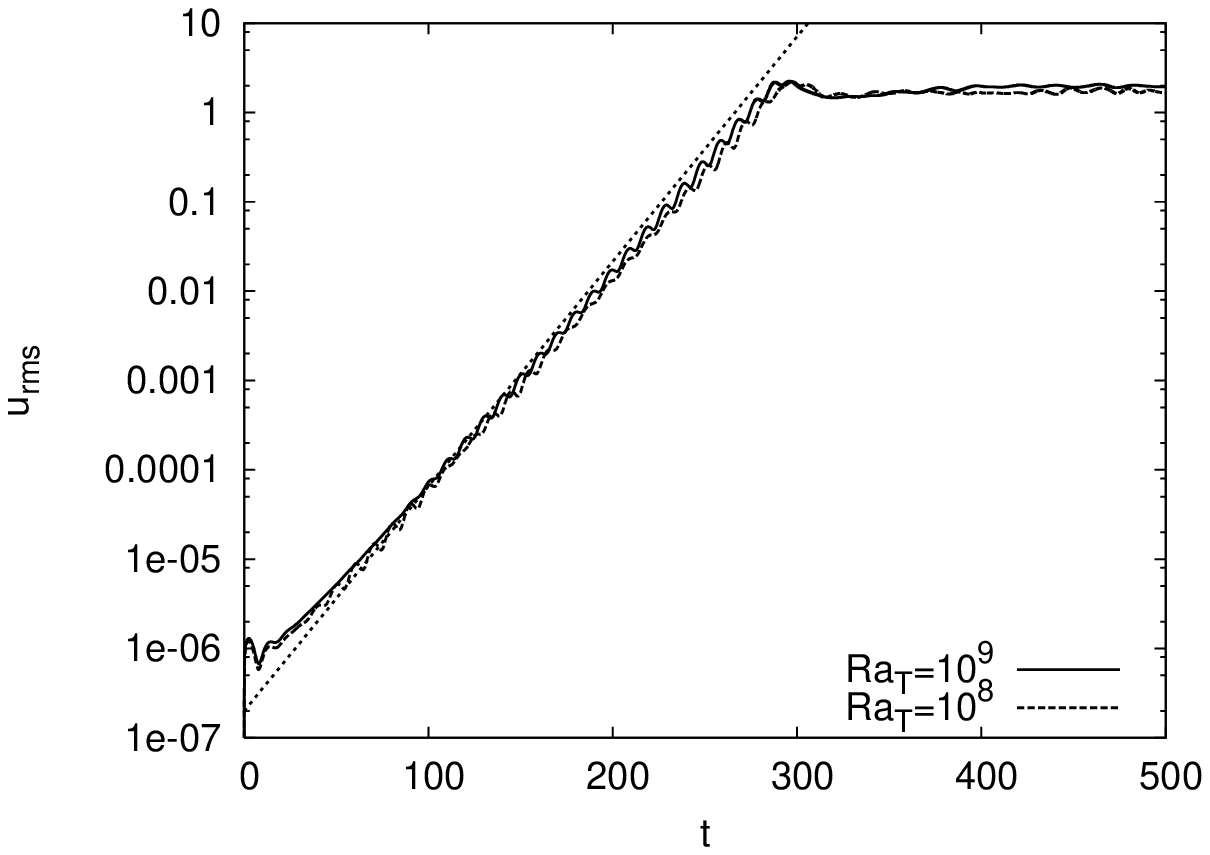}
\caption{Temporal evolution of the non-dimensional rms velocity in simulations
  with $R_0^{-1} = 1.2$, for two different computational domain
  heights: $L_z = 100$ corresponding to ${\rm Ra}_T = 10^8$ (dashed line), 
and $L_z = 178$ corresponding to ${\rm Ra}_T = 10^9$ (solid line). The straight dotted 
line shows an estimate of the early exponential growth based on the growth rate of the most
rapidly growing mode only (see \S\ref{sec:linear}). The two
simulations saturate at the same level, confirming that the dynamics
of the system in the saturated phase are independent of the domain
size (for large enough domains).} 
\label{fig:linear}
\end{figure}

The initial exponential growth is well-approximated by linear theory. This is illustrated in Figure \ref{fig:linear} for example, which shows the early temporal evolution of the rms velocity for the $R^{-1}_0 = 1.2$ run, and compares it with the growth of the fastest-growing mode according to linear theory.
The fit is very good -- the small discrepancy at early times can be attributed to the fact that more than one mode are excited, but the most rapidly growing mode then quickly takes over. We confirmed that the initial instability is independent of the Rayleigh number (i.e. the domain height) by comparing these results with those of a taller-domain simulation ($L_z = 178$) at the same $R_0^{-1}$.

\begin{figure}[h]
\epsscale{0.7}
\plotone{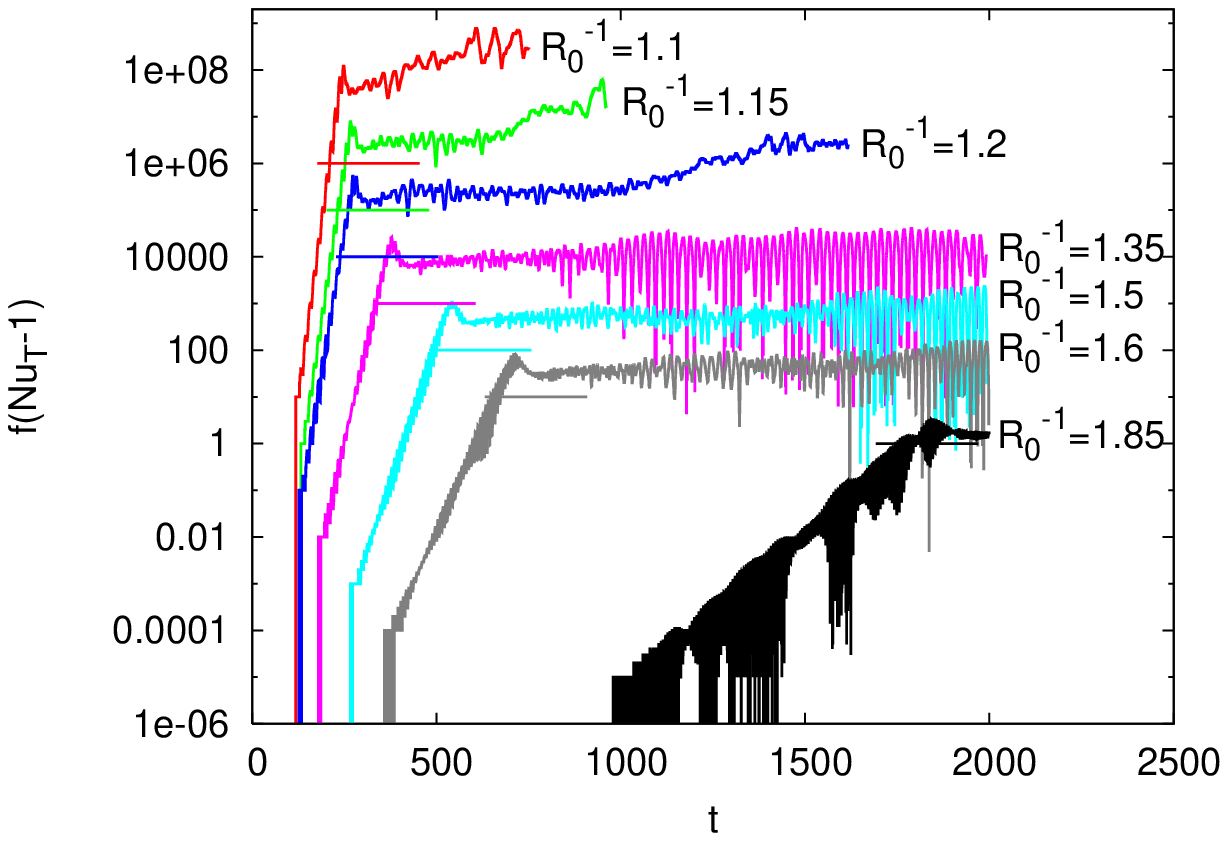}
\caption{Evolution of the thermal Nusselt number for seven of the simulations
  presented in Table 1. In all cases, ${\rm Pr}=\tau=1/3$, ${\rm Ra}_T=10^8$ ($L_z = 100$), and the
  aspect ratio is one. ${\rm Nu}_T-1$ is shown to emphasize the exponential growth phase. The results are also
  staggered for clarity, so each curve actually shows $f ({\rm Nu}_T-1)$ where the multiplicative factor $f$ is 1, 10, 100, 1000, $10^4$, $10^5$, and $10^6$ respectively for $R_0^{-1} = 1.85, 1.6, 1.5, 1.35, 1.2, 1.15$ and $1.1$. A straight horizontal line of the same color in each case marks the point at which ${\rm Nu}_T-1 = 1$ for reference (i.e. when turbulent and diffusive fluxes are equal to one another). Note how runs with $R_0^{-1} \ge 1.35$ remain in a quasi-steady saturated state, while runs with $R_0^{-1} < 1.35$ show a subsequent increase in transport. The $R_0^{-1} = 1.35$ run was actually integrated until $t = 2500$, but was found to remain at the same saturated level. } 
\label{fig:allruns}
\end{figure}

The perturbations saturate once nonlinear effects become important. 
We find that the level of saturation of the turbulence is also independent of the domain height
(alternatively, of the Rayleigh number), as shown in Figure
\ref{fig:linear}. However, it depends {\it sensitively} on the value of the density ratio. 
Figure \ref{fig:allruns} shows the temporal evolution of the thermal Nusselt number ${\rm Nu}_T$
as a function of the inverse density ratio.  A Nusselt number is the ratio of the total flux 
(diffusive + turbulent) to the diffusive flux, so we define the thermal 
Nusselt number as 
\begin{equation}
{\rm Nu}_T = \frac{-\kappa_T T_{0z} + < wT> }{-\kappa_T T_{0z} } =
1 + < \tilde{w} \tilde{T} >  \mbox{   , }
\label{eq:nutdef}
\end{equation}
where the angular brackets denote a spatial average over the entire
domain. We also define the equivalent compositional Nusselt number ${\rm Nu}_\mu$ as
\begin{equation}
{\rm Nu}_\mu = \frac{-\kappa_\mu \mu_{0z} + < w\mu> }{-\kappa_\mu \mu_{0z} }  = 1 + \frac{R_0}{\tau} < \tilde{w} \tilde{\mu} >  \mbox{   . }
\label{eq:numudef}
\end{equation} 
In each case, in the second expression the turbulent fluxes are expressed
in non-dimensional form recalling that $T_{0z} < 0$ and $\mu_{0z}<0$ 
while $T$ and $\mu$ are non-dimensionalized using $|T_{0z}|$. 

In all simulations presented in Figure \ref{fig:allruns}, the thermal Nusselt number increases exponentially until
saturation, and remains approximately constant during the early
saturated phase. After saturation, however, simulations which were run using a lower
$R_0^{-1}$ behave in a fundamentally different way from those at higher $R_0^{-1}$. In latter case,
(for $R^{-1}_0 \ge 1.35$), the Nusselt number at saturation remains statistically steady
for the entire duration of the run. By contrast, for $R^{-1}_0 < 1.35$, the Nusselt
number later continues to increase.

\begin{figure}[h]
\epsscale{1}
\plotone{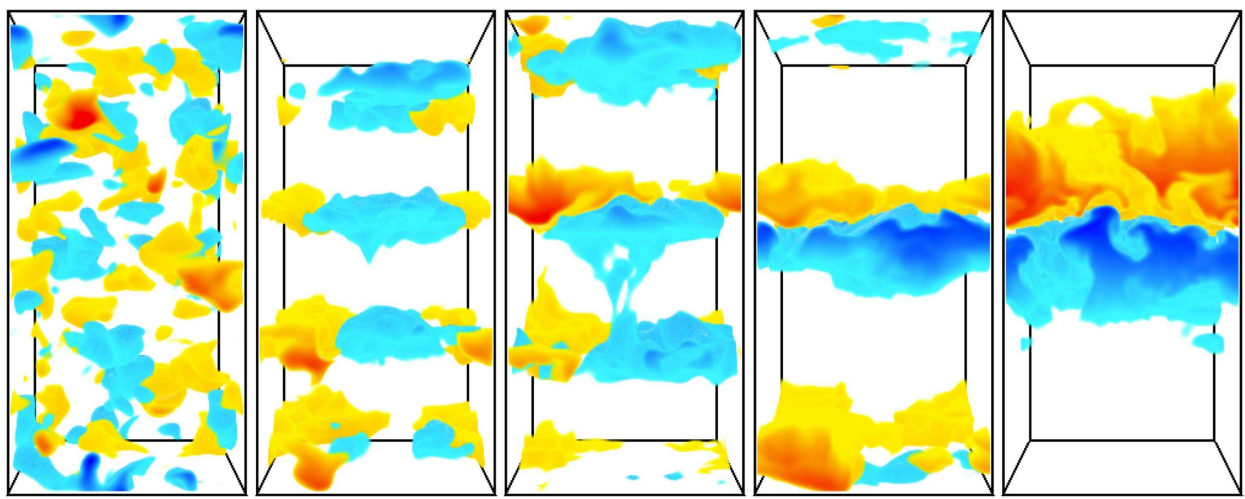}
\caption{Volume-rendered visualization of the 
  mean molecular weight perturbation, for $R_0^{-1} = 1.2$, using the
  tall-domain simulation ($L_z = 178d$). Shown
  are five snapshots taken at different times, (a) in the homogeneous
  phase at $t = 400$, (b) in the four-layer phase at $t = 1100$, (c) 
  three-layer phase at $t = 1350$, (d) two-layer phase at $t = 1550$ and (e) 
  single-layer phase at $t = 1850$. The color scale is adjusted in each panel to emphasize the perturbations, so that $\tilde{\mu} \in [-0.1,0.1] \mu_{0z} L_z$ in (a), $\tilde{\mu} \in [-0.25,0.25] \mu_{0z} L_z$ in (b)  and (c), $\tilde{\mu} \in [-0.4,0.4] \mu_{0z} L_z$ in (d)  and $\tilde{\mu} \in [-0.5,0.5] \mu_{0z} L_z$ in (e).    } 
\label{fig:layers}
\end{figure}

When visualizing the results, we find that this second increase in the turbulent transport properties of
the system corresponds to the formation of well-mixed fully convective layers separated by thin stably stratified interfaces (see Figure \ref{fig:layers} for example). The Nusselt number continues to increase as the layers merge, until a single layer is left. We have therefore established that layers can indeed form in low-Prandtl number double-diffusive 
convection, and that, as in the high-Prandtl number regime, a layered system transports heat more efficiently than a homogeneous system with the same overall temperature and compositional gradient. 
We now study both the homogeneous phase and the layered phase in more detail in \S\ref{sec:homogen} and \S\ref{sec:layers} respectively. 

\section{Homogeneous Double-Diffusive Convection}
\label{sec:homogen}
 
We focus here on the homogeneous phase, prior to the formation of the first set of layers, and measure the transport properties of the turbulence via the respective Nusselt numbers defined in (\ref{eq:nutdef}) and (\ref{eq:numudef}). 
Note that the time period between the initial saturation of the
double-diffusive instability and the onset of layer formation, when it occurs, varies
with $R_0^{-1}$ (see Figure \ref{fig:allruns}). Table 1 indicates, for
each simulation, the time interval over which the system is in this homogeneously turbulent phase and during which we average the instantaneous Nusselt numbers. 
 
\begin{figure}[h]
\epsscale{0.7}
\plotone{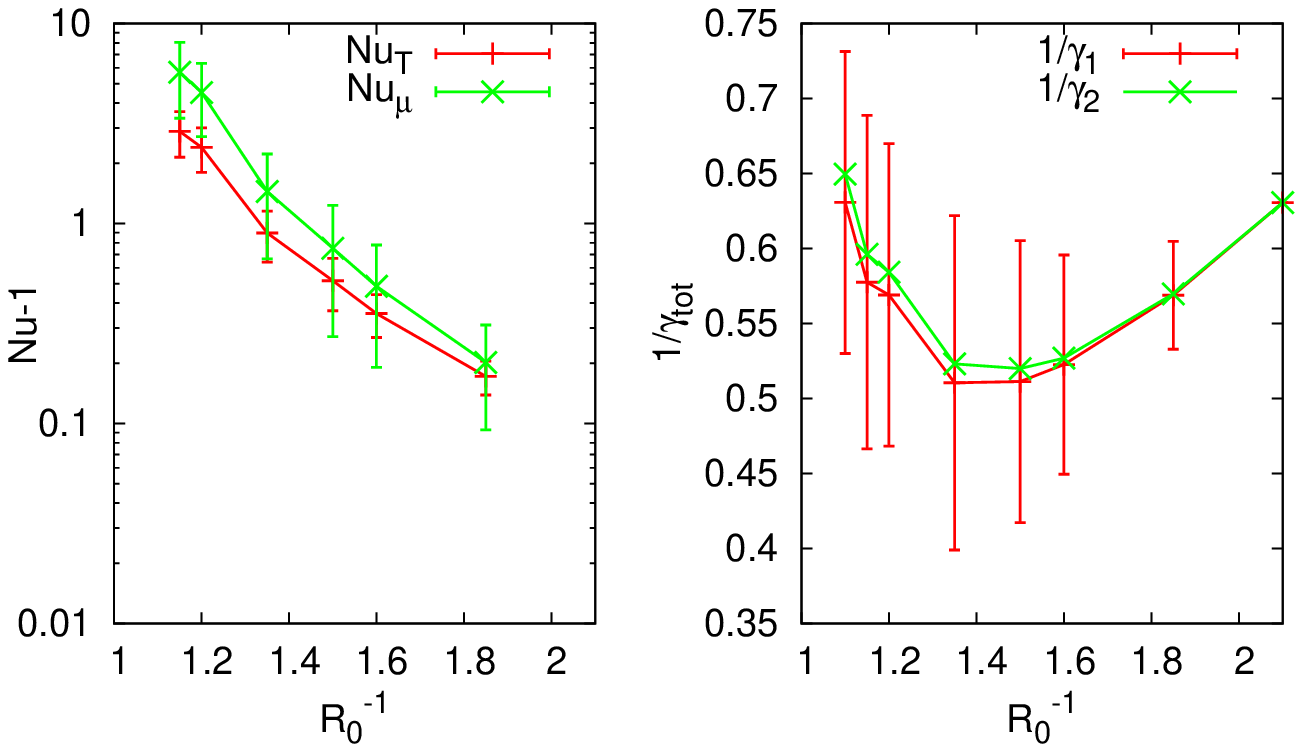}
\caption{Left: Mean Nusselt numbers in the homogeneously turbulent phase (see Table 2), as a function of the inverse density ratio. The ${\rm Nu}_T$ measurements are shown as ($+$) symbols, and the ${\rm Nu}_\mu$ measurements as $(\times)$ symbols. The turbulent contribution to transport, represented by ${\rm Nu}-1$ in both cases, goes from a few to zero over the instability range. The $R_0^{-1} = 2.1$ point was left out since we measured ${\rm Nu}_T = {\rm Nu}_\mu = 1$ in that run. Right: Inverse of the total buoyancy flux ratio $\gamma^{-1}_{\rm tot}(R_0^{-1})$ as measured in our simulations, using two different methods (see main text for detail). Note how $\gamma^{-1}_{\rm tot}$ shows a pronounced minimum around $R_0^{-1} = 1.4$. In all cases, the errorbars represent rms fluctuations of the respective functions around the mean.  } 
\label{fig:nugamma}
\end{figure}

The mean Nusselt numbers thus extracted are presented in Table 1 and illustrated in Figure
\ref{fig:nugamma}a. The errors quoted denote the rms of the fluctuations around the respective means. For most values of $R_0^{-1}$, we ran a series of simulations with different resolution, or different box size, or both. The measured Nusselt numbers are always consistent within the errorbars.

As expected, turbulent mixing is negligible close to marginal stability, i.e. when $R_0^{-1} \rightarrow ({\rm Pr} + 1)/({\rm Pr}+\tau)$. It increases as $R_0^{-1}$ decreases through the instability range, 
and grows rapidly close to the onset of overturning convection (i.e. as $R_0^{-1} \rightarrow 1$). 
However, we find that it remains fairly weak, with ${\rm Nu}_T$ of the order of a few and ${\rm Nu}_\mu$ of the order of ten, even for our lowest $R_0^{-1}$ run ($R_0^{-1} = 1.1$). By contrast, homogeneous Rayleigh-B\'enard convection in the absence of compositional gradient, in the same parameter regime (Pr = 1/3, Ra$_T
= 10^8$ and aspect ratio one), would have a thermal Nusselt number of the order of several
thousands \citep{garaudogilvie2010}. This shows that while turbulent mixing is not negligible in this
homogeneous double-diffusive regime, it is nevertheless much smaller than that induced by standard convection.   
Presumably, there exists a very narrow range of inverse density ratio close to unity, but above it, across which 
turbulent mixing rapidly but continuously increases towards the fully convective value. This will be the subject of a subsequent study. 

Since this quantity is crucial to the theory of layer formation in the fingering regime \citep{radko2003mlf,traxler2010b}, we 
also compute the so-called total flux ratio $\gamma^{\rm tot}$,
defined as the ratio of the total buoyancy flux due to heat transport
to that due to compositional transport\footnote{Note that $F^{\rm
    tot}_T$ and $F^{\rm tot}_\mu$ are dimensionless here. The
  interpretation of $\gamma^{\rm tot}$ as a buoyancy flux is more
  apparent when we go back to the dimensional quantities $F^{\rm tot, dim}_T$ and $F^{\rm tot, dim}_\mu$, since $F^{\rm tot}_T =  F^{\rm tot,dim}_T / \kappa_T |T_{0z}|$ and $F^{\rm tot}_\mu =  F^{\rm tot,dim}_\mu / (\alpha/\beta) \kappa_T |T_{0z}|$. }: 
\begin{equation}
\gamma_{\rm tot} = \frac{F^{\rm tot}_T}{F^{\rm tot}_\mu} = \frac{ R_0 }{\tau} \frac{{\rm Nu}_T}{{\rm Nu}_\mu}  \mbox{   . }
\label{eq:gammatot}
\end{equation}
There are two different ways of computing $\gamma_{\rm tot}$ from our numerical results: the first and preferable method involves calculating $\gamma_{\rm tot}(t)$ at every timestep\footnote{To be precise, we calculate $\gamma^{-1}_{\rm tot}(t)$ at every timestep, take the average of this function (shown in Figure \ref{fig:nugamma}b), and then take its inverse to get the mean $\gamma_{\rm tot}$. The reason for doing this in two steps is that there are very occasional events where ${\rm Nu}_\mu(t) < 0$, at which point $\gamma_{\rm tot}(t) \rightarrow \infty$ but the inverse remains well-defined.} and then taking its mean during the homogeneous turbulent phase. The 
second involves using the measured mean values of ${\rm Nu}_T$ and ${\rm Nu}_\mu$ directly into (\ref{eq:gammatot}). These two versions of $\gamma_{\rm tot}$ are denoted as $\gamma_1$ and $\gamma_2$
respectively, in Figure \ref{fig:nugamma}b. They are consistent within the errorbars.

Note that we are actually plotting the function $\gamma^{-1}_{\rm tot}(R_0^{-1})$ in Figure \ref{fig:nugamma}b. The reason for showing it rather than $\gamma_{\rm tot}(R_0^{-1})$ will be revealed in \S\ref{sec:gamma-instab}, and is related to the layer formation mechanism. As we shall see, the fact that $\gamma^{-1}_{\rm tot}$ is a {\it decreasing} function of $R_0^{-1}$ in the interval $[1,1.35]$, i.e. the same interval in which layers are observed to emerge, is not a coincidence. 

\section{Layered convection}
\label{sec:layers}
 
As discussed in \S\ref{sec:numdesc}, in all simulations with $ R_0^{-1} < 1.35$ we find that the system 
does not remain for long in a state of homogeneous double-diffusive convection, but spontaneously develops thermo-compositional layers instead. We study this process in more detail in this section, focusing on the
$R^{-1}_0=1.2$ run. We choose this value of $R_0^{-1}$ rather than one closer to unity, based on the results of Figure \ref{fig:allruns}. Indeed, for $R^{-1}_0=1.2$ layer formation and mergers are ``slow enough'' to be studied, but proceed much more rapidly for lower $R_0^{-1}$.  

\subsection{General considerations}
\label{sec:general-layer}

The results presented in Figure \ref{fig:allruns} are for a cubic box with height $L_z = 100$. In that run, we observe that two layers initially form, then merge into one a little while later. In order to minimize the influence of the finite domain height on the layer formation and merger process, we use from here on the taller-domain simulation, for which $L_z = 178$ (${\rm Ra}_T = 10^9$). The appearance and successive  merger of layers observed in that run is shown in Figure \ref{fig:layers}.  We find that twice as many layers initially form in this nearly-twice-as-tall domain, and appear roughly at the same time as they did in the cubic box run. 
This shows that layer formation depends only on local processes, and knows about the thermal
scale $d$ rather than the domain scale. The initial layer height, in each case, is of the order of $45d - 50d$.

\begin{figure}[h]
\epsscale{0.7}
\plotone{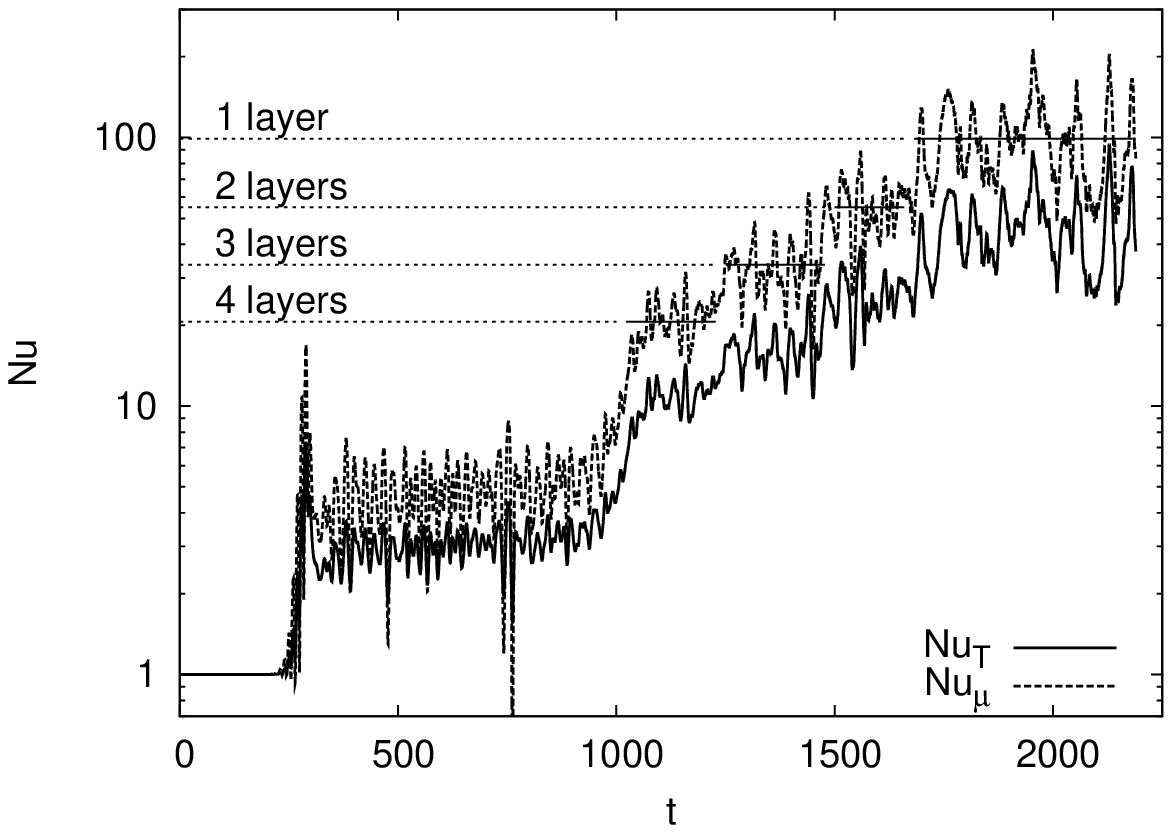}
\caption{Evolution of ${\rm Nu}_T$ (solid line) and ${\rm Nu}_\mu$ (dashed line) for the $R_0^{-1}=1.2$ tall domain run ($L_z = 178$). This plot shows the step-wise increase in transport through the various phases. In the layered phase in particular, the heat transport through the staircase depends on the layer height. The horizontal lines indicate the mean compositional Nusselt number measured in the four-, three-, two- and one-layer state in the tall-domain run  (see \S\ref{sec:fluxes} and Table 2 for detail). The solid part of each line indicates the interval of time over which the averages were measured. Equivalent lines for ${\rm Nu}_T$ are left out to avoid cluttering the plot. Note how closely the two curves follow each other throughout the entire run. } 
\label{fig:nuevol-layers}
\end{figure}

Figure \ref{fig:nuevol-layers} shows the evolution of the thermal and compositional Nusselt numbers in the tall-domain simulation.  We see quite clearly the stepwise increase in transport which 
accompanies the layer formation and successive mergers: there are five fairly well-defined plateaus, corresponding to the homogeneous phase, (up to about $t=1000$, see Figure \ref{fig:layers}a), the four-layer phase (up to about $t = 1200$, see Figure \ref{fig:layers}b), the three-layer phase (up to $t = 1450$, see Figure \ref{fig:layers}c), the two-layer phase (up to $t = 1650$, see Figure \ref{fig:layers}d), and the final one-layer phase, see Figure \ref{fig:layers}e.
It is also interesting to note how the heat and compositional fluxes follow each other closely throughout the entire simulation. 



A useful way of studying the formation and structure of the layers was
presented by \citet{stellmach2010}, and consists in looking at
Fourier modes of the density perturbation:
\begin{equation}
\tilde{\rho}(x,y,z,t) = \sum_{l,m,k} \hat \rho_{l,m,k}(t) e^{ilx+imy+ikz} \mbox{   , }
\end{equation}
where $k$ is an integer multiple of $2\pi/L_z$ and similarly for $l$ and $m$.
These Fourier modes are straightforwardly extracted from our numerical solutions since our code is spectral. 

\begin{figure}[h]
\epsscale{0.7}
\plotone{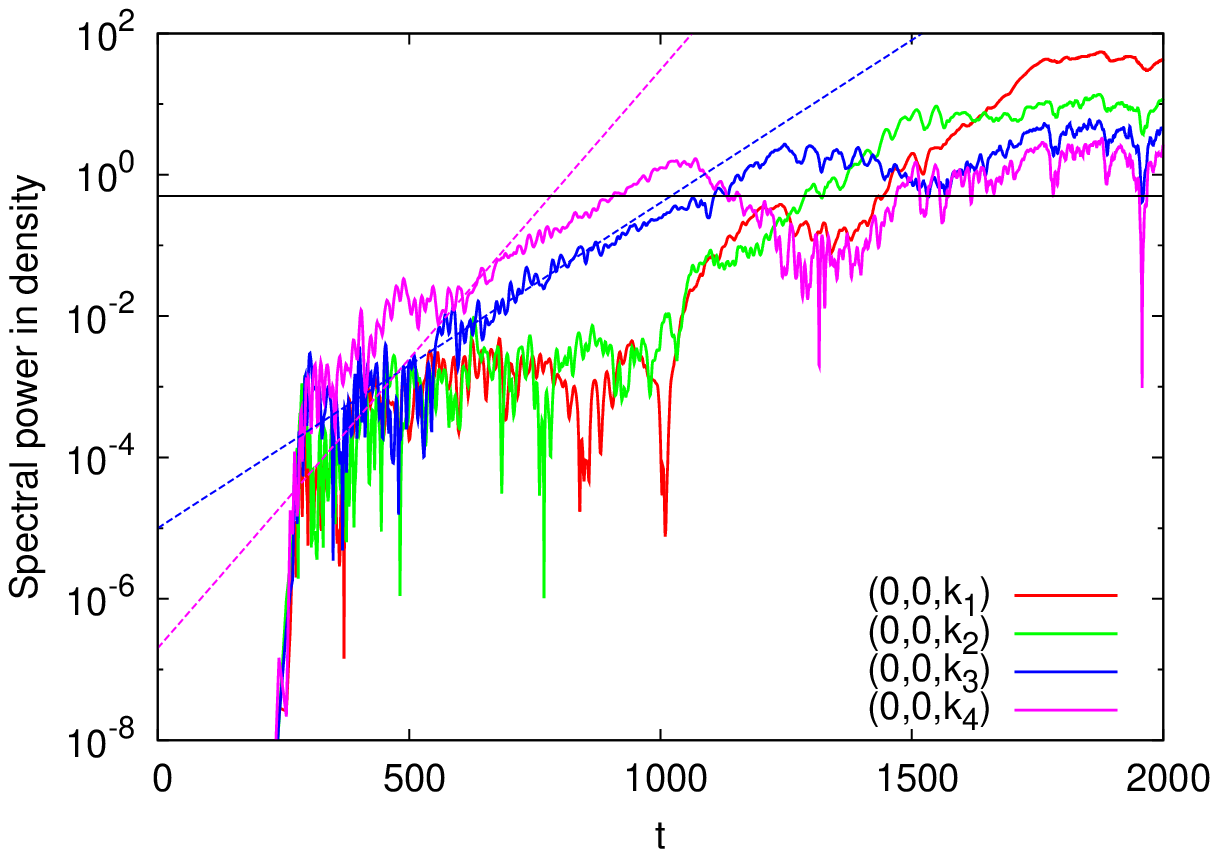}
\caption{Evolution of the norm of the Fourier coefficient of each of the four gravest Fourier modes of the vertical density perturbation profile, $|\hat \rho_{0,0,k}|^2=\hat \rho_{0,0,k}\hat \rho^*_{0,0,k}$. The figure illustrates the initial exponential growth of the $k_4$ and $k_3$ modes, and compares them with the prediction from the $\gamma-$instability theory (see \S\ref{sec:gamma-instab}), shown as the same-color straight lines. The mode grows until it reaches the critical amplitude for overturning (horizontal black line), see main text for detail. Shortly afterward, four equally spaced layers appear (around $t=1000$). The layers later merge successively, which can be seen here easily as the $k_3$, $k_2$ and $k_1$ mode respectively take over as the dominant mode. } 
\label{fig:layermodes}
\end{figure}

By definition, the $\hat \rho_{0,0,k}$ modes are the vertical Fourier modes 
of the horizontally-averaged density profile.  Since thermo-compositional staircases are also 
density staircases (i.e. with a nearly uniform density within the layers separated by sharp pycnoclines), a staircase with $n$ layers has a dominant vertical wavenumber $k_n = 2\pi n/ L_z $. This is seen most clearly in Figure \ref{fig:layermodes}, which shows the norm of $\hat \rho_{0,0,k}$ for the four gravest non-zero modes, and 
illustrates how $k_4$, $k_3$, $k_2$ and $k_1$ successively take over as the dominant mode as the layers form and merge.  

A rather striking feature of Figure \ref{fig:layermodes}, however,  is that the $k_4$ and $k_3$ modes actually begin to grow as early as $t=500$, long before the layers appear in visual inspection of the temperature and composition field (as in Figure \ref{fig:layers} for example). An even more striking result is that this growth is well-approximated by an exponential. This strongly suggests that layer formation arises through a secondary {\it linear} instability of homogeneous double-diffusive convection (see next section) rather than through stochastic overturning events of the growing gravity waves, as is commonly assumed \citep[e.g.][]{stevenson1979,spruit1992}. 

We now study in more detail, successively, the layer formation process, and the evolution of the fluxes through the staircase as the mergers proceed.

\subsection{The $\gamma-$instability}
\label{sec:gamma-instab}

\subsubsection{The $\gamma-$instability of fingering convection}

Recently, significant progress has been made in understanding the spontaneous formation of layers in {\it fingering} convection, thanks to the groundbreaking work of \citet{radko2003mlf} in the oceanographic context. Radko discovered that homogeneous fingering convection in that regime is linearly unstable to a secondary large-scale instability, which takes the form of exponentially-growing horizontally-invariant perturbations in the density (equivalently, temperature and composition) profile.  The perturbations grow  
through positive feedback between the perturbed stratification and the modulated fingering fluxes. Upon reaching a critical amplitude, the modulated density profile becomes unstable to direct overturning convection, and the system rapidly transitions into a fully-formed, regularly-spaced staircase. A sufficient condition for this instability to occur
is that the ratio of the total buoyancy flux due to heat to the total buoyancy flux due to salt, the quantity referred to as $\gamma_{\rm tot}$ in \S\ref{sec:homogen}, should be a strictly decreasing function of the density ratio. For this reason, this new instability was called the $\gamma-$instability and the associated perturbations, the $\gamma-$modes. 

Radko's theory was validated first through two-dimensional simulations \citep{radko2003mlf}, and more crucially through 3D simulations \citep{stellmach2010}. \citet{stellmach2010} were the first to find spontaneous layer formation in 3D simulations of fingering convection in the ``oceanic'' parameter regime.  They analyzed the growth and structure of the emergent layers using the vertical Fourier modes of the density profile, as we have done in the previous section. They found, as we do in Figure \ref{fig:layermodes}, that the Fourier mode which corresponds to the number of layers of the emerging staircase began to grow long before the layers form, and that its growth rate could be predicted very accurately by Radko's $\gamma-$instability theory (see their Figure 6). 

\citet{traxler2010b} extended Radko's theory for layer
formation in fingering convection to a parameter regime more relevant
of the astrophysical context. Their results suggest that at low Prandtl number and low diffusivity ratio $\gamma_{\rm tot}$ is always an increasing function of the density ratio, so that $\gamma-$modes are stable. They concluded that spontaneous layer formation is unlikely in astrophysical fingering convection. 

\subsubsection{The $\gamma-$instability of double-diffusive convection}

Radko's theory is quite general, and can be applied with only minor modifications to the case of double-diffusive convection. 
Let us consider a system in a homogeneously turbulent state, with a background density ratio $R_0$. We know through the results of \S\ref{sec:homogen} that this system drives a non-zero total vertical heat flux $F^{\rm tot}_{T}$ and a non-zero total vertical compositional flux $F^{\rm tot}_{\mu}$ with
\begin{equation}
F^{\rm tot}_{T} =  {\rm Nu}_T (R_0) \mbox{   and  } F^{\rm tot}_{\mu} = \frac{ F^{\rm tot}_{T}  }{\gamma_{\rm tot}(R_0)} \mbox{   . }
\end{equation}
As long as the background is homogeneous, however, $F^{\rm tot}_{T}$ and
$F^{\rm tot}_\mu$ are constant in the domain, and do not affect
the temperature or chemical composition of the system. 

Let us now assume that this homogeneously turbulent system is modulated by a large-scale, horizontally-invariant perturbation,  so that the horizontally-averaged temperature and mean-molecular weight profiles, $\bar T$ and $\bar \mu$ can be written as
\begin{equation}
\bar T(z,t) = \hat{T} e^{ikz + \Lambda t}\mbox{   , }
\end{equation}
and similarly for $\bar \mu$. These large-scale perturbations change the local density ratio, which we now write as $R_\rho(z,t) = R_0 + R'(z,t)$. Since the turbulent fluxes are functions of $R_\rho$, as shown in Figure \ref{fig:nugamma}a, then the perturbations also induce a spatial modulation of the turbulent fluxes. In adequate circumstances, the convergence/divergence of the modulated fluxes reinforce the original temperature and compositional perturbations, and close the feedback loop. We show in Appendix A that the growth rate of the perturbations, $\Lambda$, is the solution of the following quadratic:
\begin{equation}
\Lambda^2 + \Lambda k^2 \left[  A_2 \left( 1 -
    \frac{R_0}{\gamma_0} \right) + {\rm Nu}_0 ( 1 - A_1
  R_0) \right] - A_1 k^4 R_0 {\rm Nu}_0 = 0 \mbox{   , }
\label{eq:gamma-instab}
\end{equation}
where 
\begin{eqnarray}
&& A_1 = R_0 \left. \frac{d(1/\gamma_{\rm tot})}{dR_\rho} \right|_{R_0} \mbox{   , }
\quad A_2 = \left. R_0 \frac{d {\rm Nu}_T}{dR_\rho} \right|_{R_0} \mbox{   , } \\
&& {\rm Nu}_{0} = {\rm Nu}_T(R_0), \quad \gamma_0 = \gamma_{\rm tot}(R_0) \mbox{   . }
\end{eqnarray}
This quadratic is exactly the same as that of \citet{traxler2010b} and
by proxy that of \citet{radko2003mlf} provided his $\gamma$ is
interpreted as the total flux ratio $\gamma_{\rm tot}$, see Appendix A
for detail. As discussed by \citet{radko2003mlf} there is a positive
real root when $A_1 > 0$, i.e. when $\gamma_{\rm tot}$ decreases with
increasing density ratio, or alternatively, when $\gamma^{-1}_{\rm tot}$
decreases with $R_0^{-1}$. Exactly as in the case of fingering convection, {\it a necessary 
condition for layer formation in double-diffusive convection 
is that $\gamma_{\rm tot}$ should be a decreasing function of $R_0$}. 

Based on the results of Figure \ref{fig:nugamma}b, we can now
straightforwardly explain, thanks to this theory, the dichotomy between the cases with
$R_0^{-1} > 1.35$, for which we do not expect (and do not observe)
layer formation, and the cases with $R_0^{-1} < 1.35$ for which we
do. The case $R_0^{-1} = 1.35$ is unclear given our measurement errors
on $\gamma_{\rm tot}$. Furthermore, since the mode growth rate increases with
$A_1$, we can also explain, at least qualitatively, why
the staircase forms much more rapidly at lower $R_0^{-1}$ (i.e. because the $\gamma_{\rm tot}$ curve is steeper in that regime). In what follows, we now compare theory and simulations more
quantitatively.

\subsection{Comparison of the $\gamma-$instability theory with numerical results} 

The tall-domain simulation described in \S\ref{sec:general-layer} shows the emergence of a four-layer staircase. 
Based on the discussion of the $\gamma-$instability of \S\ref{sec:gamma-instab}, we need to compare the growth rate of the $k_4$ mode shown in Figure \ref{fig:layermodes} to the solution of (\ref{eq:gamma-instab}) with $k = k_4$. In order to do this, we first estimate the coefficients $A_1$ and $A_2$. The derivatives of Nu$_T$ and $\gamma_{\rm tot}$ with respect to $R_\rho$, at $R_\rho = R_0 = \frac{1}{1.2}$, are calculated using the fluxes obtained in simulations at neighboring values of the inverse density ratio. We find that
\begin{equation}
A_1 =  0.453 \mbox{   , } A_2 =  12.9 \mbox{   . }
\end{equation}
We also use the value of ${\rm Nu}_0$  (and associated errorbar) given in Table 1 for $R_0^{-1} = 1.2$, and the average of the two $\gamma_{\rm tot}$ values for $\gamma_0$. 

We find that the dominant $\gamma-$mode has a dimensionless growth rate
\begin{equation}
\Lambda(k_4) = 9.42 \times 10^{-3} \mbox{   . }
\end{equation}
Similarly, we find that 
\begin{equation}
\Lambda(k_3) = \Lambda(k_4) \frac{k^2_3}{k_4^2} = 5.30 \times 10^{-3}  \mbox{   . }
\end{equation}
Figure \ref{fig:layermodes} compares these theoretical predictions with the numerical results for the $k_4$ and $k_3$ modes. We find that they over-predict the growth rate of the dominant $k_4$ mode, but provide a fairly accurate estimate of the growth rate of the $k_3$ mode.

The fact that the $k_4$ mode grows somewhat slower than predicted is actually expected, and leads us to discuss an important caveat of the $\gamma-$instability theory. The derivation of (\ref{eq:gamma-instab}) is fundamentally based on an assumption of scale separation between the small-scale turbulence, which drives the heat and compositional fluxes, and the large-scale temperature and compositional perturbations $\bar T$ and $\bar \mu$. However, solutions of (\ref{eq:gamma-instab}) satisfy the similarity law $\Lambda \propto k^2$, implying that for any given $\gamma-$mode there exists, in theory, another more rapidly growing one with smaller wavelength. This ``ultra-violet catastrophe'' problem was discussed by \citet{radko2003mlf}, and is clearly an artefact of the model. In practice, the $\gamma-$instability theory should only be applied for $k$ significantly smaller than the wavenumber of the fastest growing mode of the primary instability, in order to satisfy the required separation of scales. 

It is therefore interesting to see that the mode which emerges as the dominant $\gamma-$mode in Figure \ref{fig:layermodes} has a wavelength of about $45d$, which is only about twice as large as the wavelength of the most rapidly growing primary gravity wave (about $20d$, see \S\ref{sec:linear}). As such, it appears to be the smallest-scale mode for which the $\gamma-$instability theory remains applicable, since it still grows exponentially, albeit with  a growth rate somewhat slower than predicted. This behavior is very similar to the one found by  \citet{stellmach2010} in the case of fingering convection. It is reassuring to see that, by comparison, the larger-scale $k_3$ mode grows as well, and this time with a growth rate which is much closer to the predicted one. 

\subsection{Staircase formation}

\begin{figure}[h]
\epsscale{0.7}
\plotone{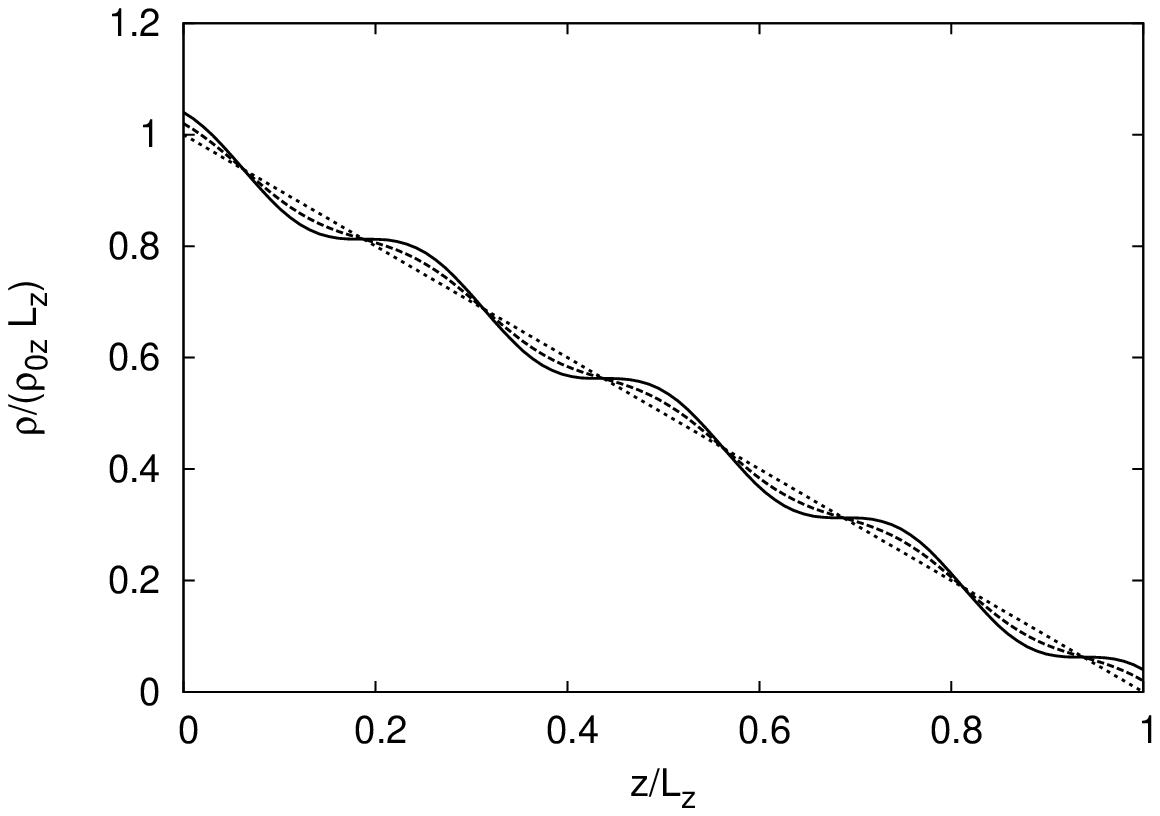}
\caption{Evolution of the density profile of a $k_4$ $\gamma-$mode. For simplicity, the density is normalized to the density difference across the domain height, and the vertical coordinate is in units of the domain height. The dotted line shows the background density profile with constant gradient $\rho_{0z}$, the dashed line the background + perturbation for intermediate perturbation amplitude, while the solid line shows the background + perturbation at the critical amplitude for the onset of overturning convection given by (\ref{eq:rhocrit}). Note the existence of four specific points where the total density gradient $\rho_{0z} + d\rho/dz$ is  zero, which will become the middle of the emergent layers.} 
\label{fig:gamma-turn}
\end{figure}

The dominant $\gamma-$mode grows in amplitude until it causes regularly spaced local 
inversions in the density profile (see Figure \ref{fig:gamma-turn}). The critical 
amplitude $\hat{\rho}_{\rm crit}$, for which the total density gradient $ \rho_{0z}  + d \rho/dz = 0$ 
somewhere in the domain, depends 
on the mode wavenumber and on the background density ratio \citep[e.g.][]{stellmach2010}: 
\begin{equation} 
\hat{\rho}_{\rm crit} = \frac{1-R_0^{-1}}{k} \mbox{   . }
\label{eq:rhocrit}
\end{equation}

As the mode grows beyond this amplitude (shown as a horizontal line in Figure \ref{fig:layermodes}), progressively larger regions of the domain are unstable to direct overturning convection, and a fully-formed staircase rapidly appears. 

Once layers have formed, the $\gamma-$instability theory no longer applies and the mode stops growing. 
The subsequent evolution of the staircase through mergers is caused by subtle differences in the fluxes through the interfaces and through the convective regions. While a full study of the merger dynamics is beyond the scope of this paper, we present in the next section an analysis of the heat and compositional fluxes though the staircase as a function of the layer height. 

\subsection{Layer mergers and flux increase}
\label{sec:fluxes}

The time of each merger roughly corresponds, in Figure \ref{fig:layermodes}, to the point where the amplitude of the $k_{n}$ mode overtakes that of the $k_{n+1}$ mode. Shortly afterward, the new dominant mode stops growing, at
which point the staircase has reached a new temporary $n-$layered equilibrium.
We can then measure the transport properties of the system 
while it is in this $n-$layered state. The results are 
presented in Table 2 and illustrated in Figure \ref{fig:nuevol-layers}. 

\begin{figure}[h]
\epsscale{0.7}
\plotone{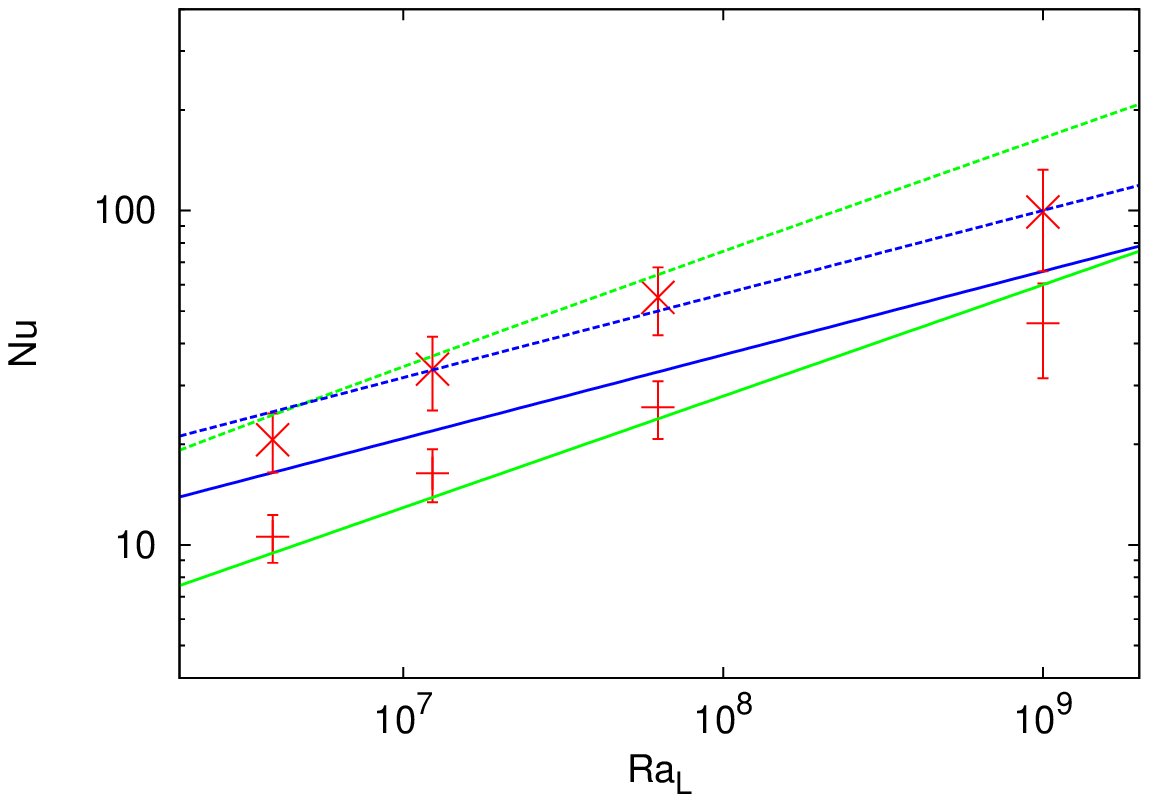}
\caption{Variation of the two Nusselt numbers in the layered phase, for the tall-domain $R_0^{-1} = 1.2$ run, with the Rayleigh number defined  with layer height, ${\rm Ra}_L$. The symbols correspond to the measurements presented in Table 2, with ${\rm Nu}_T$ shown as ($+$) symbols and  ${\rm Nu}_\mu$ shown as ($\times$) symbols. 
The errorbars show the standard deviation of the measured fluxes about the mean. The various lines correspond to different possible scaling laws discussed in the main text: the blue lines are from \citet{spruit1992}, for ${\rm Nu}_T$ (solid line) and ${\rm Nu}_\mu$ (dashed line), and the green lines are from (\ref{eq:convection-from-wall}) for ${\rm Nu}_T$ (solid line) and (\ref{eq:numuconvection-from-wall}) for ${\rm Nu}_\mu$ (dashed line). } 
\label{fig:Nuvslayers}
\end{figure}

Assuming that the sum of the thicknesses of all the interfaces is small compared with the height of the domain, we deduce the layer height $H_L(n) = L_z/n = 178d/n$. We can then construct a Rayleigh number based on the layer height rather than the domain height, 
\begin{equation}
{\rm Ra}_L = \left(\frac{H_L}{d}\right)^4 \mbox{   . }
\label{eq:Ralayers}
\end{equation}
Figure \ref{fig:Nuvslayers} shows the variation of ${\rm Nu}_T$ with ${\rm Ra}_L$.
Rather remarkably, we find that within the errorbars the heat flux through the system is more or less consistent with the standard scaling law for convection between two bounded plates separated by a distance $H_L$ \citep[see][and references therein]{garaudogilvie2010},
\begin{equation}
{\rm Nu}_T \simeq 0.06 {\rm Ra}_L^{1/3} \mbox{   . }
\label{eq:convection-from-wall}
\end{equation}
Note that no fitting was involved here. This scaling would be consistent with an interpretation of the interfaces as impermeable, diffusive boundary layers.

The compositional Nusselt number is about twice the thermal Nusselt number, in the layered phase and in the homogeneous phase. A possible explanation for this result is the following. The {\it turbulent} flux ratio $\gamma = <\tilde{w} \tilde{T}>/<\tilde{w}\tilde{\mu}>$ is typically of order unity for turbulence induced by double-diffusive instabilities\footnote{A plausible reason for this is the following, as argued by \citet{radko2003mlf}. As $R_0^{-1} \rightarrow 1$, the compositional field acts more and more like a passive tracer. As a result, the turbulent diffusivities for heat and composition tend to one-another, and since $R_0 = 1$, so do the induced turbulent buoyancy fluxes. As $R_0^{-1} \rightarrow ({\rm Pr}+\tau)/({\rm Pr}+1)$, $\gamma$ tends to one for a different reason. Since the instability is driven by the conversion of potential energy into kinetic energy, near the marginal stability limit this available potential energy must vanish. Hence, the turbulent buoyancy flux due to heat must exactly equal that due to composition for the potential energy available to vanish. Finally, since $\gamma$ has to tend to one in both limits, it cannot deviate significantly away from one inbetween.} \citep{radko2003mlf}. As a result, using (\ref{eq:nutdef}) and (\ref{eq:numudef}) we have:
\begin{equation}
{\rm Nu}_\mu \simeq 1 + \frac{R_0}{\tau} <\tilde{w} \tilde{T}> = 1 + \frac{R_0}{\tau} \left( {\rm Nu}_T - 1\right) = 1 + 2.5 \left( {\rm Nu}_T - 1\right) 
\label{eq:numuconvection-from-wall}
\end{equation}
for our selected parameters. For large enough Nusselt numbers, this implies ${\rm Nu}_\mu \simeq 2.5 {\rm Nu}_T$.
Figure \ref{fig:Nuvslayers} shows (\ref{eq:numuconvection-from-wall}) in comparison with the data. Again, the fit is satisfactory within the errorbars, although not quite as compelling as that of ${\rm Nu}_T$. Note that a similar argument was invoked by \citet{traxler2010b} to explain the relationship between their measured scaling laws for ${\rm Nu}_T$ and ${\rm Nu}_\mu$ in fingering convection.

It is also interesting to compare our numerical results with the work of \citet{spruit1992}, who proposed the following parametrization for heat transport by layered convection: 
\begin{equation}
{\rm Nu}_T \simeq 0.5 ({\rm Pr} {\rm Ra}_L)^{1/4} \mbox{   . }
\label{eq:spruit92}
\end{equation}
His estimate of the turbulent compositional diffusivity (see his equation (44)) yields the following compositional Nusselt number:
\begin{equation}
{\rm Nu}_\mu \propto \tau^{-1/2} \frac{\nabla}{\nabla_\mu} {\rm Nu}_T = R_0  \tau^{-1/2} {\rm Nu}_T 
\end{equation}
in this Boussinesq case, where the proportionality constant is of order unity. Both estimates are shown in Figure \ref{fig:Nuvslayers} as well, and are also more-or-less consistent with our flux measurements within the errorbars.
This time, ${\rm Nu}_\mu$ seems to be better accounted for than ${\rm Nu}_T$. 

Further simulations will be needed to determine which (if any) of these scaling laws best explains the data. Wider domains will be needed to improve our statistics and reduce the variability of the mean fluxes. In addition, taller domains will be necessary to allow for larger layer heights. But beyond the details of the power laws or the prefactors, our simulations clearly show that the overall transport through a staircase depends only on the layer height, and that the heat and compositional fluxes are roughly proportional to each other for given $R_0^{-1}$, ${\rm Pr}$ and $\tau$. 


\begin{table}
\begin{center}
\caption{Variation of the Nusselt numbers as a function of the number of layers, for the $R_0^{-1}=1.2$, $L_z = 178d$ run.}
\begin{tabular}{cccccc}
\\
\tableline
$ n $ & $H_L $ & $t_{\rm start}$   & $t_{\rm end}$      & ${\rm Nu}_T$    & ${\rm Nu}_\mu$       \\ 
4        & 44.5        &1020      & 1240        & 10.5 $ \pm$ 1.7            & 20.6 $\pm$ 4.1      \\
3        & 59.3        &1240       & 1480        & 16.4$ \pm$ 2.9            & 33.5 $\pm$ 8.4      \\
2        & 88.9          & 1480      & 1660        & 25.8 $\pm $ 5.0              & 55.0 $\pm$ 12.6 \\
1        & 177.8         &1660       & 2190        & 46.0 $\pm $ 14.5              &  99.0 $\pm$ 33.3  \\
 \tableline
\end{tabular}
\tablecomments{The first column shows the number of layers, and the second 
column an estimate of the layer depth in units of $d$, $H_L= L_z / n$. The times $t_{\rm start}$ and $t_{\rm end}$ mark
  the interval of the time over which the turbulent fluxes, measured
  via ${\rm Nu}_T$ and ${\rm Nu}_{\mu}$, were averaged. Errorbars represent the rms fluctuations about the mean.}
\end{center}
\end{table}

\section{Discussion and conclusion}
\label{sec:ccl}

\subsection{Summary of the results}

In this work, we have studied a set of numerical simulations of double-diffusive convection, in a triply-periodic domain, for Prandtl number ${\rm Pr} = \nu/\kappa_T = 1/3$ and diffusivity ratio $ \tau = \kappa_\mu/\kappa_T = 1/3$. We have explored the entire instability range, varying the inverse density ratio $R_0^{-1}$ between 1 (the onset of direct overturning convection) and $({\rm Pr} + 1)/({\rm Pr} + \tau)$ (the marginal stability limit). Our simulations were 
performed in a ``small'' domain spanning, in the horizontal direction, about five wavelengths of the fastest-growing double-diffusive mode (i.e. $L_x = L_y = 100d$ where $d$ is a thermal diffusion lengthscale), and in the vertical direction, $L_z = 100d$ or $L_z = 178d$ depending on the runs. 

In all cases we initialized a double-diffusively unstable system with infinitesimal perturbations, and found that these  first grow exponentially according to linear theory, then saturate into a state of homogeneous double-diffusive convection. In that state, the turbulent contribution to thermal and compositional transport is significant but much smaller than that expected from standard convection, ranging from 5-10 times the diffusive rate near the onset of direct convective instability, and rapidly dropping towards zero as $R_0^{-1}$ increases towards marginal stability (see \S\ref{sec:homogen}). 

For small $R_0^{-1} $, however, the system does not remain in this homogeneously convecting state. Instead, thermo-compositional layers rapidly appear, and transport through the system strongly increases. We showed 
that the layer formation process is governed by Radko's $\gamma-$instability theory \citep{radko2003mlf,stellmach2010,traxler2010b},  both qualitatively and quantitatively. In particular, it explains why our simulations with $R_0^{-1} <1.35$ transition into layers while those with $R_0^{-1} > 1.35$ do not. The key factor is the variation of the total buoyancy flux ratio $\gamma_{\rm tot}$ with density ratio (see \S\ref{sec:gamma-instab}): layers can only form when $\gamma_{\rm tot}$ decreases with $R_0$. 

In the layered phase, we found that the flux through the staircase depends sensitively on the mean layer 
height $H_L$. Given the large variability of the measured fluxes during the layered phase, our results are roughly consistent both with Spruit's theory \citep{spruit1992} and with heat transport between two solid plates (as given by equation (\ref{eq:convection-from-wall})). Further simulations will be needed to help distinguish between these two possibilities -- or perhaps suggest an alternative one. Finally, note that in our small-domain simulations, the mergers always proceed until a single layer 
is left. In that sense, the dynamical evolution of the system is always eventually influenced by the domain size. 

\subsection{Discussion of the applicability of our results to real systems}

Our initial goals were threefold: (a) to characterize transport by homogeneous double-diffusive convection (i.e. in the absence of layers), (b) to determine if, under which conditions, and through which process thermo-compositional layers may form and (c) to characterize transport by layered double-diffusive convection when appropriate. 

To answer part (a) in detail, a much larger number of simulations will be needed, using progressively smaller Prandtl numbers and diffusivity ratios. These are the subject of an ongoing investigation. We hope to find similar scaling laws for ${\rm Nu}_T$ and ${\rm Nu}_\mu$ as functions of {\rm Pr} and $\tau$ as the one found by \citet{traxler2010} for fingering convection. 

By contrast with the case of fingering convection, however, we now know that thermo-compositional staircases 
{\it can} form spontaneously from double-diffusive convection. Radko's criterion for layer formation, namely that $\gamma_{\rm tot}$ should decrease with $R_0$, is the answer to part (b) of our goals, but does require knowledge of the function $\gamma_{\rm tot}(R_\rho)$ to be applied in practice. The latter must be determined separately for each parameter set $({\rm Pr}, \tau)$. 

Finally, our findings have provided some insight into part (c). Since transport through a staircase depends on the layer height only (for given fluid parameters and overall stratification), the problem shifts to estimating actual layer heights in astrophysical objects. The layers we observe in our simulations have a strong tendency to merge, which suggests two possible outcomes: the mergers continue indefinitely, until the scale of the equilibrium layers is commensurate with the system size; or the mergers stop for other reasons (see below), with an equilibrium layer height significantly smaller than the system size. It is of course crucial to know which of these two scenarios is correct, as they imply vastly different transport rates through the staircase. 

Unfortunately, our simulations were not able to provide a definitive answer to this question. In the two cases studied, the final layer height was equal to the domain height, but this should not be interpreted as a result in favor of the first scenario since this could simply mean that our domain was too small to ``contain'' the intrinsic equilibrium layer height of the second scenario. 

\citet{radko2005dtl} proposed a theory supporting the idea that mergers stop before layers reach the system size, and deduced a means of estimating the equilibrium layer height. Starting from an initial staircase with uniform ``jumps'' in temperature and chemical composition across the interfaces,  he studied how the staircase evolves if it is perturbed slightly, by making some of the jumps larger and some of the jumps weaker. He concluded that the staircase is unstable to mergers if the total buoyancy flux ratio through the interfaces is a decreasing function of the density ratio across the interfaces -- a criterion very similar to the $\gamma-$instability criterion. 

We have tried to test Radko's merger theory against our simulations, but this has unfortunately proven to be difficult. The statistical fluctuations in the measured turbulent fluxes (see Figure \ref{fig:nuevol-layers} for example) are too large to detect a significant variation of the interfacial flux ratio as the mergers proceed. We would need a much larger domain to improve the signal to ``noise'' ratio to a point where our results could be compared with his theory. We would also need a much taller domain (at least a few times taller than the equilibrium layer height) to see if the merger process indeed stops as expected. Finally, we would need to integrate the simulation long enough to establish convincingly that the mergers have indeed stopped. Unfortunately, running equivalent simulations in a much larger domain, and for long enough to observe the layer formation and merger process, is impossible within current numerical limitations. 

\subsection{Future prospects}

The preliminary findings presented in this paper still enable us to lay out a clear path towards obtaining better parametrizations of mixing by double-diffusive convection in the near future, using currently available computational resources:
\begin{itemize}
\item Firstly, we must gain a better understanding of the instability saturation mechanism at low Prandtl number and low diffusivity ratio, in order to determine the flux laws ${\rm Nu}_T({\rm Pr}, \tau, R_0)$ and ${\rm Nu}_\mu({\rm Pr}, \tau, R_0)$ for homogeneous double-diffusive convection. These flux laws are needed to determine when layers are expected to form, and can be used ``as is'' to parametrize mixing otherwise. They {\it can} be measured using ``small-domain'' simulations similar to the ones we have presented here, at least for values of ${\rm Pr}$ and $\tau$ as low as about 0.01 or so. Semi-analytical weakly-nonlinear models will then be helpful to guide extrapolations to the much lower parameter values appropriate of the astrophysical regime. 
\item Secondly, we must gain a general understanding of mixing in the layered case, at low Prandtl number and low diffusivity ratio. In order to do this, we need to determine how interfacial transport depends on the fluid parameters $({\rm Pr},\tau)$ and on the interfacial density ratio (ie. a density ratio based on the difference in temperature and composition across the layers). We must also understand how transport scales {\it within} the convective layers, as a function of the same parameters but also as a function of the layer height. This can be done today using simulations in which a single layer is pre-seeded, to bypass the rather lengthy layer formation and merger phases. Using this information, we will be able to test the basic flux laws which are central to Radko's merger theory more quantitatively \citep{radko2005dtl}. If this theory holds, then one can straightforwardly deduce the equilibrium layer height for a given parameter set, and ultimately quantify the staircase transport properties.
\end{itemize}

\acknowledgments

E.R, P.G.  and A.T are supported by funding from the NSF (NSF-0807672). 
S.S. was supported by grants from the NASA Solar and
Heliospheric Program (NNG05GG69G, NNG06GD44G, NNX07A2749). All
computations were performed on the UCSC Pleiades supercomputer,
purchased with an NSF-MRI grant. 

\appendix

\section{Derivation of the $\gamma-$instability in the diffusive case. }

Following \citet{radko2003mlf} and \citet{traxler2010b}, we begin with the general non-dimensional governing equations (\ref{eq:goveqs}), and average them over several wavelengths of the fastest growing mode of the primary instability. We get 
\begin{eqnarray}
\frac{1}{\mathrm{Pr}} \left(\frac{\partial\mathbf{u}}{\partial t} + \mathbf{u}\cdot\nabla \mathbf{u} \right) & = & -\nabla p + (T - \mu){\bf e}_z + \nabla^2 \mathbf{u} - \frac{1}{\mathrm{Pr}}\nabla \cdot \mathbf{R} \label{eq:mean_momentum}, \\
\frac{\partial T}{\partial t} - w + \mathbf{u}\cdot\nabla { T}
& = & - \nabla \cdot
\mathbf{F}^{\rm tot}_T, \label{eq:mean_temperature} \\
\frac{\partial \mu}{\partial t} - \frac{1}{R_0}w + {\bf u}\cdot\nabla \mu & = & -  \nabla \cdot \mathbf{F}^{\rm tot}_\mu \label{eq:mean_composition},
\end{eqnarray}
where $\mathbf{R}$ is the Reynolds stress, $\mathbf{F}^{\rm tot}_T$ and $\mathbf{F}^{\rm tot}_\mu$ are the total heat and compositional fluxes respectively, and $T$, $\mu$, $p$, and $\bu$ now denote large-scale fields only. 

The $\gamma-$instability drives horizontally-invariant perturbation with zero mean flow \citep{radko2003mlf}. 
We can therefore neglect the momentum equation, set ${\bf u} = 0$, and ignore all horizontal
derivatives. The mean temperature and composition equations simplify to: 
\begin{eqnarray}
\frac{\partial T}{\partial t}  & = & - \frac{\partial F_T^{\rm    tot}}{\partial z} , \nonumber \\
\frac{\partial \mu}{\partial t} & = & - \frac{\partial F_\mu^{\rm
    tot}}{\partial z}    \mbox{   . }
\end{eqnarray}

Finally, we assume that ${\rm Nu}_T$, and $\gamma_{\rm tot}$ depend
only on the local value of the density ratio $R_\rho$.  Note that $R_\rho$ is no longer constant, but varies with $z$ as a result of the large-scale background temperature and compositional perturbations, as
\begin{equation}
R_\rho = \frac{\alpha (T_{0z} + T^{\rm dim}_z)}{\beta ( \mu_{0z} + \mu^{\rm dim}_z) } = \frac{R_0  (1 - T_z)}{ 1 - R_0 \mu_z }  \mbox{   , }
\label{eq:Rrhodef}
\end{equation} 
where, for clarity, we first expressed $R_\rho$ as the ratio of dimensional quantities and then as the ratio of non-dimensional quantities.

We now linearize equations (\ref{eq:mean_temperature}) and (\ref{eq:mean_composition}) around a state of homogeneous turbulent convection in which $T = 0 + T'$, $\mu = 0 + \mu'$, and $R_\rho = R_0 + R'$ where linearization of (\ref{eq:Rrhodef}) yields:
\begin{equation}
R' = R_0(1 - T_z + R_0 \mu_z) \mbox{   . }
\label{R0_linear}
\end{equation}
Using the fact that $F^{\rm tot}_T = {\rm Nu}_T(R_\rho) (1-T_z)$ and $F^{\rm tot}_\mu = F^{\rm tot}_T /\gamma_{\rm tot}$ 
the linearized temperature equation becomes
\begin{equation}
\frac{\partial T'}{\partial t}  = A_2 \left(\frac{\partial^2 T'}{\partial z^2} - R_0 \frac{\partial^2 \mu'}{\partial z^2}\right) + {\rm Nu}_0\, \frac{\partial^2 T'}{\partial z^2} \mbox{   ,}
\label{T_linear}
\end{equation}
while the linearized composition equation is
\begin{equation}
\frac{\partial \mu'}{\partial t} =\frac{1}{\gamma_0} \frac{\partial T'}{\partial t} +   A_1 \left(\frac{\partial^2 T'}{\partial z^2} - R_0  {\rm Nu}_0  \frac{\partial^2 \mu'}{\partial z^2}\right)   \mbox{   , }
\end{equation}
where we have used the following notation for simplicity:
\[
\begin{array}{ll}
A_1 = R_0 \left. \frac{d(1/\gamma_{\rm tot})}{dR_\rho} \right|_{R_0}, & A_2 = \left. R_0 \frac{d {\rm Nu}_T}{dR_\rho} \right|_{R_0}, \\
{\rm Nu}_{0} = {\rm Nu}_T(R_0), & \gamma_0= \gamma_{\rm tot}(R_0).
\end{array}
\]
Assuming normal modes of the form $T', \mu' \sim e^{i k  z + \Lambda t}$,
we get the quadratic
\begin{equation}
\Lambda^2 + \Lambda k^2 \left[  A_2 \left( 1 -
    \frac{R_0}{\gamma_{\rm tot}} \right) + {\rm Nu}_0 ( 1 - A_1
  R_0) \right] - A_1 k^4 R_0 {\rm Nu}_0 = 0 \mbox{   . }
\end{equation}
This quadratic is exactly the same as the one obtained in the fingering case. In hindsight, this 
result is trivial, and could be obtained immediately had we allowed ourselves to non-dimensionalize
$T$ and $\mu$ using negative dimensions (in which case the governing equations and all 
definitions are exactly the same as in the fingering case).


\end{document}